\documentclass[letterpaper,twocolumn,10pt]{article}
\usepackage{usenix2019_v3}

\usepackage{tikz}
\usepackage[normalem]{ulem}
\usetikzlibrary{tikzmark}

\usepackage{multirow}
\usepackage{makecell}
\usepackage{adjustbox}
\usepackage{array}

\usepackage{hyperref}
\usepackage{url}
\usepackage{amsmath}
\usepackage{amsfonts}
\usepackage{mathtools}
\usepackage{xspace}
\usepackage{booktabs}
\usepackage{multirow}
\usepackage{graphicx}
\usepackage{grffile}
\usepackage{longtable}
\usepackage{xcolor, soul}
\usepackage{subfig}
\usepackage{caption}
\def\arraystretch{1}
\usepackage{pifont} 
\usepackage{xcolor,colortbl}
\usepackage{pdflscape}
\usepackage{enumitem}
\newcommand{\tick}{\ding{51}}
\newcommand{\cross}{\ding{55}}
\usepackage{cleveref}

\usepackage{comment}

\usepackage{titlesec}
\usepackage{lipsum}

\usepackage{filecontents}

\newcommand{\PreserveBackslash}[1]{\let\temp=\\#1\let\\=\temp}
\newcolumntype{C}[1]{>{\PreserveBackslash\centering}p{#1}}

\newcommand{\system}{\ensuremath{\mathsf{Hermes}}\xspace}

\newcommand{\ntce}{\ensuremath{\mathtt{NEUTREX}}\xspace}
\newcommand{\securoberta}{\ensuremath{\mathtt{CellulaRoBERTa}}\xspace}
\newcommand{\irsynthesizer}{\ensuremath{\mathtt{IRSynthesizer}}\xspace}
\newcommand{\keyextractor}{\ensuremath{\mathtt{Keyword~Extractor}}\xspace}
\newcommand{\astgenerator}{\ensuremath{\mathtt{DTGenerator}}\xspace}
\newcommand{\irtranslator}{\ensuremath{\mathtt{IRTranslator}}\xspace}
\newcommand{\fsmsynthesizer}{\ensuremath{\mathtt{FSMSynthesizer}}\xspace}

\newcommand{\uegen}{\ensuremath{\mathtt{UEGEN}}\xspace}

\newcommand{\trancomp}{\ensuremath{\mathsf{TC^{NL}}}\xspace}
\newcommand{\trancompplural}{\ensuremath{\mathsf{TC^{NL}}}s\xspace}
\newcommand{\trancompir}{\ensuremath{\mathsf{TC^{IR}}}\xspace}

\newcommand{\newattacks}{3\xspace}
\newcommand{\prevattacks}{19\xspace}
\newcommand{\attacksfp}{8\xspace}

\newcommand{\numdevices}{9\xspace}
\newcommand{\numdeviations}{7\xspace}
\newcommand{\deviationsfp}{3\xspace}

\newcommand{\numdsl}{94\xspace}

\newcommand\blfootnote[1]{%
  \begingroup
  \renewcommand\thefootnote{}\footnote{#1}%
  \addtocounter{footnote}{-1}%
  \endgroup
}

\newcommand{\rrcsetupreq}{\scriptsize\textsc{RRCSetupRequest}\normalsize\xspace}
\newcommand{\rrcsetup}{\scriptsize\textsc{RRCSetup}\normalsize\xspace}

\newcommand{\rrcresumereq}{\scriptsize\textsc{RRCResumeRequest}\normalsize\xspace}
\newcommand{\rrcresume}{\scriptsize\textsc{RRCResume}\normalsize\xspace}

\newcommand{\rrcrelease}{\scriptsize\textsc{RRCRelease}\normalsize\xspace}
\newcommand{\rrcreject}{\scriptsize\textsc{RRCReject}\normalsize\xspace}

\newcommand{\rrcconreconfig}{\scriptsize\textsc{RRCConnectionReconfiguration}\normalsize\xspace}

\newcommand{\ueinforeq}{\scriptsize\textsc{UEInformationRequest}\normalsize\xspace}
\newcommand{\ueinforesponse}{\scriptsize\textsc{UEInformationResponse}\normalsize\xspace}

\newcommand{\measreport}{\scriptsize\textsc{MeasurementRReport}\normalsize\xspace}

\newcommand{\attachreq}{\scriptsize\textsc{ATTACH REQUEST}\normalsize\xspace}
\newcommand{\attachaccept}{\scriptsize\textsc{ATTACH ACCEPT}\normalsize\xspace}

\newcommand{\attachreject}{\scriptsize\textsc{ATTACH REJECT}\normalsize\xspace}

\newcommand{\servicereq}{\scriptsize\textsc{SERVICE REQUEST}\normalsize\xspace}

\newcommand{\servicereject}{\scriptsize\textsc{SERVICE REJECT}\normalsize\xspace}

\newcommand{\identityreq}{\scriptsize\textsc{IDENTITY REQUEST}\normalsize\xspace}
\newcommand{\identityresponse}{\scriptsize\textsc{IDENTITY RESPONSE}\normalsize\xspace}

\newcommand{\authreq}{\scriptsize\textsc{AUTHENTICATION REQUEST}\normalsize\xspace}
\newcommand{\authresponse}{\scriptsize\textsc{AUTHENTICATION RESPONSE}\normalsize\xspace}
\newcommand{\authfailure}{\scriptsize\textsc{AUTHENTICATION FAILURE}\normalsize\xspace}
\newcommand{\authreject}{\scriptsize\textsc{AUTHENTICATION REJECT}\normalsize\xspace}

\newcommand{\smcommand}{\scriptsize\textsc{SECURITY MODE COMMAND}\normalsize\xspace}
\newcommand{\smcomplete}{\scriptsize\textsc{SECURITY MODE COMPLETE}\normalsize\xspace}
\newcommand{\smreject}{\scriptsize\textsc{SECURITY MODE REJECT}\normalsize\xspace}

\newcommand{\emminfo}{\scriptsize\textsc{EMM-INFORMATION}\normalsize\xspace}

\newcommand{\gutireallocationcommand}{\scriptsize\textsc{GUTI REALLOCATION COMMAND}\normalsize\xspace}
\newcommand{\gutireallocationcomplete}{\scriptsize\textsc{GUTI REALLOCATION COMPLETE}\normalsize\xspace}

\newcommand{\detachreq}{\scriptsize\textsc{DETACH REQUEST}\normalsize\xspace}

\newcommand{\regreq}{\scriptsize\textsc{REGISTRATION REQUEST}\normalsize\xspace}

\newcommand{\regreject}{\scriptsize\textsc{REGISTRATION REJECT}\normalsize\xspace}

\newcommand{\deregreq}{\scriptsize\textsc{DEREGISTRATION REQUEST}\normalsize\xspace}

\newcommand{\ctlblock}{\footnotesize\texttt{ctl\_block}\normalsize\xspace}

\newcommand{\startstatetag}{\footnotesize\texttt{start\_state\_tag}\normalsize\xspace}
\newcommand{\enstatetag}{\footnotesize\texttt{end\_state\_tag}\normalsize\xspace}   
\newcommand{\condtag}{\footnotesize\texttt{cond\_tag}\normalsize\xspace}

\newcommand{\acttag}{\footnotesize\texttt{act\_tag}\normalsize\xspace}

\newcommand{\token}{\footnotesize\texttt{token}\normalsize\xspace}

\newcommand{\errclr}[1]{\textcolor{red}{#1}}

\begin{document}

\date{}

\title{\Large \bf \system: Unlocking Security Analysis of Cellular Network Protocols by Synthesizing Finite State Machines from Natural Language Specifications 
}

\author{
{\rm Abdullah Al Ishtiaq, Sarkar Snigdha Sarathi Das, Syed Md Mukit Rashid, Ali Ranjbar} \\
{\rm Kai Tu, Tianwei Wu, Zhezheng Song, Weixuan Wang, Mujtahid Akon}\\
{\rm Rui Zhang, Syed Rafiul Hussain}\\
Pennsylvania State University
}

\maketitle

\begin{abstract}
In this paper, we present \system, 
an end-to-end framework to automatically generate formal representations from natural language cellular specifications. 
We first develop
a neural constituency parser, \ntce, to process transition-relevant texts 
and extract transition components (i.e., states, conditions, and actions). 
We also design
a domain-specific language to translate 
these transition components to 
logical formulas
by leveraging dependency parse trees.
Finally, 
we 
compile these logical formulas to generate transitions and create the 
formal model as finite state machines.
To demonstrate the effectiveness of \system, we evaluate it on 4G NAS, 5G NAS, and 5G RRC specifications 
and obtain an overall accuracy of 81-87\%, 
which is a substantial improvement over the state-of-the-art.
Our security analysis of the extracted models uncovers \newattacks 
new vulnerabilities and identifies \prevattacks previous attacks in 4G and 5G specifications, and 
\numdeviations deviations in commercial 4G basebands.
\end{abstract}

\blfootnote{Accepted at USENIX Security '24}

\section{Introduction}

The cellular standard body, i.e., the 3rd Generation Partnership Project (3GPP), designs, creates and maintains technical specifications for cellular network systems used by network operators and equipment vendors worldwide~\cite{3gpp}.
However, a wide range of cellular stakeholders and network entities, a vast diversity of use cases such as SMS, data access, and roaming connections, and tighter backward compatibility requirements make it difficult for 3GPP to maintain the specifications in a simple form. 
Especially, the natural language description of the cellular network design spreading over hundreds of documents makes it extremely laborious and often error-prone to create formal models necessary for security analysis. 
Moreover, the inherent ambiguities and complexities of natural language often lead to misinterpretations by the developers, 
resulting in deviations and 
exploitable flaws in implementations~\cite{ltefuzz, doltest, dikeue, basespec}. 
Such deviations are difficult to identify without access to the baseline/gold standard.  
However, the cellular standard body does not provide any formal model of the system, 
thereby leaving the lack of a gold standard.

Although formal analysis of cellular network design 
has uncovered several new vulnerabilities, all of them rely on hand-crafted models~\cite{formal-5g-auth, formal-5g-aka, formal-5g-handover, formal-5g-key, lteinspector, 5greasoner}. 
Unfortunately, 
manually constructing such formal models is tedious and error-prone
due to the sheer size of cellular systems and the large number of documents.
Moreover, these models require significant time and effort 
and are often limited by modeling flaws (incorrectly representing the system) \cite{lteinspector}, inconsistent levels of abstraction \cite{5greasoner, formal-5g-auth}, and oversimplified representations of complex protocol details \cite{5greasoner, lteinspector}, leading to inadequate formal security analysis. 
Furthermore, although
3GPP introduces new generations (e.g., 3G, 4G, 5G) of cellular technology roughly every decade, 
within a generation, specifications are frequently updated, 
e.g., on average, the specification documents are updated 5-6 times each year, 
and each update includes several hundreds of line changes (details in Appendix \ref{appendix:spec-update}).
Despite not being 
major technology shifts, 
these changes in the protocol can often lead to new vulnerabilities. 
For example, changes in the collision resolution mechanism between two particular procedures introduced in the 4G specification~\cite{4g_nas_twelve} led to a new vulnerability~\cite{bookworm-game}. 
Likewise, in this work, we find a new vulnerability in 5G 
(\S\ref{sec:new-attacks}), stemming from a new cause of message rejections introduced in the 5G specification update~\cite{5g_nas_sixteen}. 
Conversely, 
due to time and effort constraints, 
frequently updating hand-crafted models is not always feasible. 
In fact, these prior models 
have never undergone revisions to account for new changes as they have emerged. 
Thus, automated extraction of formal models from cellular specifications is 
pivotal for analyzing the security of cellular protocols.

Recently, researchers used Natural Language Processing (NLP) tools to identify \textit{hazard indicators} in cellular specifications for creating concrete test cases \cite{bookworm-game} and to discover \textit{security-relevant change requests}~\cite{sr-cr}. 
Unfortunately, none of them extract formal models automatically. 
Furthermore, among the attempts to extract structured representation from natural language in other domains, e.g., protocols~\cite{rfcnlp}, geographical questions~\cite{learning-zettlemoyer, transferring-semantic-parsing, language2logic-attn}, medicine~\cite{transformation-bolc}, 
and database queries~\cite{online-zettlemoyer, transferring-semantic-parsing, language2logic-attn, spider}, 
RFCNLP \cite{rfcnlp} extracts Finite State Machines (FSMs) from Request for Comments (RFC) documents, albeit for small protocols. In contrast, cellular protocols have a substantially large number of transitions, a myriad of use cases, variables, events, message fields, and complex interactions across layers and sub-layers, rendering RFCNLP ineffective.
The absence of standardized datasets or benchmarks further compounds the difficulty in automatically extracting formal 
representations.

In this paper, we address these challenges by developing \system 
(Hermes was the God of translator/interpreter in Greek mythology) that
automatically extracts the formal representation of cellular 
network protocols from the given natural language specification documents. 
We develop three main components:  
a neural constituency parser called \ntce, \irsynthesizer, and \fsmsynthesizer. 
With these components, \system 
generates FSMs 
as a set of transitions 
containing states, conditions (e.g., received messages), and actions 
(e.g., sending messages) in a logical format. 
At a high level, 
\system first uses \ntce, 
developed with a domain-knowledge-informed grammar and deep-learning-based neural parsing model,  
to extract the portions of natural language texts related to transitions
and detect components (e.g., states, conditions, actions) within them, along with their logical relations.
\system then leverages \irsynthesizer, 
designed with a Domain Specific Language (DSL) and a neural dependency parser, 
to identify the key information items, 
including variables and events, and convert the conditions and actions  
to 
logical formulas or \emph{intermediate representations (IR)}, 
which can be easily transpiled to specific formal language 
amenable for formal analysis. 
Finally, the \fsmsynthesizer of \system combines the 
logical formulas corresponding to the 
transition components 
to construct complete transitions and the FSMs.

To demonstrate the efficacy of \system 
across multiple cellular layers, generations, and releases, 
we evaluate it on 4G NAS \cite{4g_nas}, 5G NAS \cite{5g_nas}, 5G RRC \cite{5g_rrc}, and RFC documents~\cite{rfcnlp}.
Our results show that \system-generated FSMs achieve up to 87.21\% accurate 
transitions compared to human-annotated transitions for 4G and 5G NAS specifications. 
Regarding identifying transition constituents from cellular specification, \ntce 
achieves 68.69\% labeled F1-score, 
showcasing a substantial improvement over the state-of-the-art RFCNLP's 38.52\%.
We also evaluate \system on RFCs, and it achieves 57.06\% labeled F1-score compared to 47.76\% of RFCNLP. 

To demonstrate that \system facilitates security analysis of cellular networks, 
we use it to perform two different security analyses of the cellular protocols. First, we transpile the generated formal models to SMV language~\cite{smv} by applying prior tools~\cite{lteinspector, 5greasoner} and perform model checking. 
This analysis identifies \textbf{\prevattacks} previous and \textbf{\newattacks} new vulnerabilities in the 4G and 5G specifications.  
Second, we compare the \system-generated FSM with commercial 4G cellular baseband FSMs to identify flaws in the implementations.  
This analysis uncovers \textbf{\numdeviations} deviations in \numdevices basebands with an accuracy of 78\%.

\noindent\textbf{Responsible Disclosure.} 
We have shared our findings with the GSMA CVD panel, and they have confirmed and acknowledged our contributions with CVD-2023-0071. 

\noindent\textbf{Contributions.} We summarize our contributions as follows:
\begin{itemize} [noitemsep,topsep=0pt,leftmargin=0.4cm]
    \item We design and implement \system, a framework for automatically extracting formal models from natural language cellular specification documents. We show the effectiveness of \system across different cellular generations (4G and 5G) and layers (NAS and RRC), significantly outperforming state-of-the-art in both cellular and other domains.
    \item We demonstrate the use of the formal models extracted by \system in the security analysis of the \emph{design} of 4G and 5G cellular network protocols. We identify \prevattacks previous and \newattacks new issues in those design documents. 
    \item We also show that the extracted formal representation of the 4G NAS specification effectively identifies \numdeviations \emph{implementation-level} flaws in \numdevices cellular devices by comparing those devices' FSMs with \system-generated FSM. 
    \item We provide a curated dataset 
    of natural language transition components in cellular specifications and 
    the source code of \system,  
    serving as a benchmark for future research. The dataset, source code, and properties will be available at: \url{https://github.com/SyNSec-den/hermes-spec-to-fsm}.
\end{itemize} 
\section{Background}

\noindent\ding{111} \textbf{Cellular Network Architecture.}
The User Equipment (UE), which includes USIM and cryptographic keys, is the device users use to access cellular services. 
A base station, gNB in 5G or eNodeB in 4G, provides cellular services to users and manages radio resources in a geographical cell area. 
Multiple base stations communicate with one another through a radio access network (RAN). 
The core network consists of several components or network functions (NF). 
The AMF (MME in 4G) is a major NF that is responsible for UEs' mobility management, e.g., registration, in the network through the Non-Access Stratum (NAS) protocol.

\begin{figure}[t!]
    \centering
    \begin{minipage}{0.62\linewidth}
        \centering
        \includegraphics[width=0.85\linewidth]{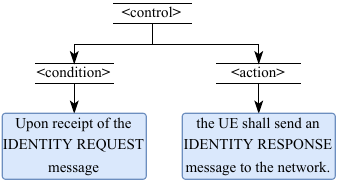}
		\caption{A constituency parse tree.}
		\label{fig:example_const}
    \end{minipage}
    \begin{minipage}{0.03\linewidth}
        \centering
    \end{minipage}
    \begin{minipage}{0.35\linewidth}
        \centering
        \includegraphics[width=0.85\textwidth]{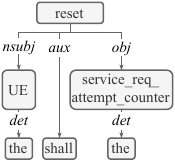}
		\caption{A dependency parse tree.
  }
		\label{fig:dep_example_tree}
    \end{minipage}
\end{figure}

\begin{figure*}[t!]
	\begin{center}
		\includegraphics[width=\textwidth]{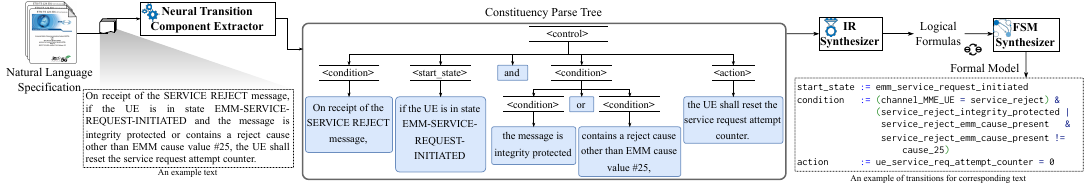}
		\caption{Overview of \system.}
		\label{fig:workflow}
	\end{center}
\end{figure*}

\noindent\ding{111} \textbf{Word Embeddings and Pretrained Language Models.} 
Word embedding represents a word as a real-valued vector, which encodes its meaning.
BERT~\cite{bert} leverages Transformer~\cite{transformer} based encoders to learn contextualized word embeddings by pretraining over general-domain corpora. 
However, Liu et al.~\cite{roberta} reveal that
the next sentence prediction objective of BERT degrades the 
embeddings, 
and the same words are masked in different training epochs of BERT, 
leading to sub-optimal training.
They incorporated these decisions into RoBERTa to deliver stronger performances.

\noindent\ding{111} \textbf{Constituency Parsing.} 
Constituency parsing builds a hierarchical tree, which describes the internal structure of a sentence, including \emph{non-terminal} constituents and \emph{terminal} input words.
Figure~\ref{fig:example_const} shows the input sentence parsed into its constituents-- condition and action spans. 
Conventional constituency parsing methods \cite{henderson2003inducing, socher2013parsing} rely on explicit grammar production rules demanding extensive domain knowledge without leveraging any constituent semantics. 
Neural constituency parsers~\cite{zhang2020fast, neural-constituency-3, kitaev2018multilingual} address this problem by learning those \textit{grammar}s during training through marginal probabilities,
which are utilized during inference. 

\noindent\ding{111} \textbf{Dependency Parsing.} While constituency parsing detects the constituent spans from a sentence, dependency parsing \cite{stanford-dependency-parser} provides the explicit relations that we need to synthesize FSMs from cellular specifications.
For example, in Figure~\ref{fig:dep_example_tree}, if a span is detected as an \textit{action} (``reset''),
we also need to know the \textbf{subject} (``UE'') and \textbf{object} (``service\_req attempt\_counter'') associated with it.
Accordingly, dependency parsing identifies the main \textbf{verb} word as the root node, and all other words have exactly one headword as their parent node. This gives a unique path from the root verb node to all other word nodes to facilitate the synthesis of logical expression.

\section{Overview of \system}
\label{sec:overview}
In this section, we define the problem, present an overview of \system, and discuss the challenges and insights.

\subsection{Problem Statement}
\label{sec:problem_statement}
Given a cellular system specification in natural language, \system aims to generate a formal model, $\mathcal{M} = \langle \mathbb{P}, \mathbb{F}, \mathbb{H}, \mathbb{V} \rangle$: 
\begin{itemize} [noitemsep,topsep=0pt,leftmargin=0.4cm]
	\item $\mathbb{P}$ is the set of protocol participants, $\{\mathcal{P}_1, \mathcal{P}_2, ...\}$, in the given specification.
	\item $\mathbb{F}$ is the set of synchronous communicating FSMs, $\{\mathcal{F}_1, \mathcal{F}_2, ...\}$, for each $\mathcal{P} \in \mathbb{P}$.
	\item $\mathbb{H}$ is the set of one-way communication channels between the 
 entities in $\mathbb{P}$, i.e., 
    $\{\mathcal{H}_{ij} | \forall \mathcal{P}_i, \mathcal{P}_j \in \mathbb{P} \land \mathcal{P}_i \neq \mathcal{P}_j \}$.
	\item $\mathbb{V}$ is the set of 
	variables, $\{\mathcal{V}_1, \mathcal{V}_2, ...\}$, 
    in $\mathcal{M}$.
\end{itemize}
We define each FSM in $\mathbb{F}$ as $\mathcal{F} = \langle \mathbb{S}, \mathcal{S}_0, \mathbb{T} \rangle$, where $\mathbb{S}$ is the set of states 
$\{\mathcal{S}_1, \mathcal{S}_2, ...\}$, 
$\mathcal{S}_0$ is the initial state, and $\mathbb{T}$ is the set of transitions. 
We further define each transition in $\mathbb{T}$ as $\mathcal{T} = \langle \mathcal{S}_i, \mathcal{S}_f, \mathcal{C}, \mathbb{A} \rangle$, where $\mathcal{S}_i$ is the starting state, and 
$\mathcal{S}_f$ is the ending state
of the transition $\mathcal{T}$, 
$\mathcal{C}$ is the logical condition 
for the transition $\mathcal{T}$ to take place, 
and $\mathbb{A}$ is the set of actions, $\{\mathcal{A}_1, \mathcal{A}_2, ...\}$, which also occur when the transition $\mathcal{T}$ takes place. 
Here, an action $\mathcal{A} \in\mathbb{A}$ can be a message transmission 
through a channel $\mathcal{H} \in \mathbb{H}$ 
or an update to a variable $\mathcal{V} \in \mathbb{V}$.

\subsection{Solution Sketch of \system}
We address the problem in \S\ref{sec:problem_statement} with three primary components of \system: Neural Transition Component Extractor (\ntce), Intermediate Representation Synthesizer (\irsynthesizer), and Finite State Machine Synthesizer (\fsmsynthesizer). The workflow is shown in Figure \ref{fig:workflow}.

\ntce takes natural language texts and identifies the constituent spans (transition components in natural language,  \trancompplural) 
corresponding to the states, conditions, actions, and the transition itself (\ctlblock). For that, we extend a state-of-the-art neural constituency parser \cite{zhang2020fast} and generate constituency parse trees for input natural language texts. 
Figure \ref{fig:workflow} demonstrates \ntce extracted \trancompplural (start\_state, action, condition, and control) and logical relations (e.g., and, or) in the constituency parse tree corresponding to the input text.

In the next stage, \irsynthesizer takes \trancompplural (state, condition, or action) within the \ctlblock, identified as constituent spans by \ntce, and constructs corresponding transition components in intermediate representations (IR), \trancompir, in a logical format. 
Finally, the generated IRs are compiled with \fsmsynthesizer to create the transitions, ultimately constituting the formal representation of the cellular specification. 

\subsection{Challenges and Insights of Extracting FSM}

\noindent\textbf{\uline{C1: 
Complex associations of FSM 
transition components in natural language.}} 
Identifying transition components (\trancompplural), i.e.,
start state, end state, conditions, and actions of an FSM's  
transitions 
from natural language specifications is challenging. This is because the associations among the components are highly complex and depend on both syntactic features and semantic relations.
For example, consider the sentence ``\textit{The UE initiates attach procedure} by sending an \attachreq to the network'' in TS 24.301~\cite{4g_nas}. 
It can have two interpretations:
(1) ``the UE shall initiate the attach procedure'', and ``the UE shall send an \attachreq to the network'', resulting in two actions by a UE without any condition,
or (2) The first part is treated as ``when the UE initiates attach procedure'' 
and thus can be considered as a condition for the second part--``sending the \attachreq message". Therefore, domain knowledge is required to disambiguate such cases of complex associations.

\noindent{\ding{109} \textbf{Insight (I1)}:} 
We adopt a domain-knowledge-informed grammar for transition components and a neural model to extract them automatically. 
First, we define a nested grammar for \trancompplural considering complex and nested associations prevalent in cellular specifications. We curate an expert-annotated dataset by labeling \trancompplural in cellular specifications accordingly. Next, we design a neural constituency parsing-based transition component extractor (\ntce) that extracts complex nested structures by comprehending semantic information. 
The resulting parse tree (as in the parse tree in Figure \ref{fig:workflow}) precisely maintains the hierarchy of different constituent spans (\trancompplural), allowing us to automatically handle the nested grammar for cellular documents. In contrast, existing neural models for network protocols \cite{rfcnlp} fail to automatically detect and handle such complex associations and nested constituents. 

\begin{figure}[t!]
	\begin{center}
		\includegraphics[width=0.95\linewidth]{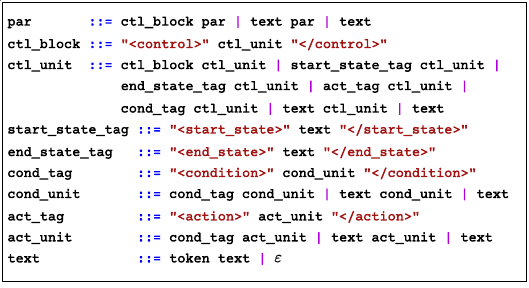}
		\caption{Defined grammar for annotated data.}
		\label{fig:grammar}
	\end{center}
\end{figure}

\noindent\textbf{\uline{C2: Limitations of existing NLP models on cellular specification.}} 
Off-the-shelf constituency parsers face several challenges when extracting cellular \trancomp. 
As cellular specifications are written in a highly technical language with 3GPP-specific abbreviations and terms,
existing word embeddings trained on general-purpose corpora are inadequate owing to the domain shift.
Also, 
existing constituency parsers do not perform reasonably well 
in the cellular dataset because of complex nested associations. 
Finally, transformer-based embedding models often face token limits surpassed by cellular document paragraphs. This requires appropriate measures to avoid truncated parse trees and inaccurate FSMs.

\noindent{\ding{109} \textbf{Insight (I2)}:
To close the domain gap in the embedding model, we curate and clean a large natural language dataset consisting of 22,000 cellular documents, including technical specifications, change requests, and technical reports, 
for unsupervised pretraining of a RoBERTa \cite{roberta} embedding model.
Second, while training the constituency parser \cite{zhang2020fast}, we observe that it struggles to extract both span boundaries and labels simultaneously in the first few steps.
Therefore, at the beginning of the training, we prioritize labeling the tokens and gradually give the same priority to identifying spans as the training progresses. Lastly, long datapoints are handled by splitting and generating embeddings for all input tokens.

\label{sec:challenge_3}
\noindent\textbf{\uline{C3: Absence of annotated cellular datasets for translation from NL phrases to logical form.}} 
\label{sec:challenge_nl_to_logic}
Transition components (\trancompplural) 
are not directly usable for any formal security analysis and 
need to be converted to logical expressions and statements, such as 
$\mathsf{UE\_service\_req\_attempt\_counter\!}$ $\mathsf{:=\!0}$.  
Existing data-driven approaches for such translation \cite{learning-iyer, semantic-parsing-coarse, semantic-type-constraints, exploring-logic} are largely domain-specific and fail to generalize across domains or capture complex logical relations~\cite{spider}. Further, no dataset exists for translating natural language to logical formulas for cellular systems,  
and annotating such a large dataset is tedious and error-prone.

\noindent{\ding{109} \textbf{Insight (I3)}:}
To circumvent the lack of annotated datasets, 
we simplify the NL-to-logic task 
and take an intermediary step similar to schema-based translations \cite{schema-1, schema-2}.
Accordingly, 
we define a Domain Specific Language (DSL) (partially shown in Figure \ref{fig:dsl}) that enables parsing and translating \trancompplural to precise and consistent logical representations.
Moreover, the DSL postulates a structure for the translation and 
highlights the required information for each logical formula through commands and corresponding arguments.
If any information is missing in \trancomp due to natural language ambiguities or constituency parsing inaccuracies, 
the DSL can guide the search for them through metadata, e.g., type information.

\label{sec:challenge_4}
\noindent\textbf{\uline{C4: Text to DSL rule mapping requires explicit arguments.}}
The DSL provides a pathway to logical formulas from natural language strings. 
However, it has a strict syntax and fixed arguments, whereas natural language is inherently diverse and imprecise. 
For example, different sentence structures may express the same meaning, such as ``the MME shall send an \attachaccept message to the UE'', 
``an \attachaccept shall be sent to the UE by the MME'', and so on. 
Mapping such diverse texts to the same DSL rule is non-trivial. 

\noindent{\ding{109} \textbf{Insight (I4)}:} 
The DSL commands are inherently tree-structured and can be represented as Abstract Syntax Trees (AST). 
Conversely,
natural languages follow sequential patterns. To map them to DSL, a tree format is a reasonable intermediate step. 
Among the different natural language tree representations,
dependency trees extract useful relations among the tokens 
and often demonstrate similar structures as the DSL ASTs. 
Figures \ref{fig:dep_ast_trees} (a) and \ref{fig:dep_ast_trees} (c) show such similarities.
Here, 
the primary verb--``receive'', is the root in both trees, and the messages are their descendants.  
Even in cases where a similar meaning is represented by different sentence structures (e.g., examples in C4), 
these 
relations are preserved reasonably well.
These similarities indicate that dependency parsing on the transition components (\trancomp) can help in mapping texts to DSL rules and, thereafter, generating logical formulas.

\section{Extracting Transition Components}
\label{ref:subsec_constituency}

The neural transition component extractor (\ntce) takes each paragraph of a cellular specification as input and 
extracts the constituency parse trees based on the grammar in Figure \ref{fig:grammar}. It identifies \ctlblock with transition components (\trancompplural) 
and simultaneously refrains from including natural language texts that do not contain any transition information. 

\subsection{Grammar and Annotated Dataset}
\label{sec:grammar-details}

As discussed in \S\ref{sec:challenge_nl_to_logic}, the \trancompplural are interrelated through complex associations and require domain expertise for correct interpretations.
To address this challenge, 
we define a grammar for \trancompplural
as shown in Figure~\ref{fig:grammar}
and annotate three 
specification documents, namely 4G-NAS-Release16 \cite{4g_nas},  
5G-NAS-Release17 \cite{5g_nas}, 
and 5G-RRC-Release17 \cite{5g_rrc}.
The annotation is done by manually inspecting each paragraph and sentence, 
and identifying states, conditions, and actions in them. 
Further, we 
consider their semantic relations and dependencies, and
follow the defined grammatical structure during annotation. 

A string 
in the language defined by this grammar can consist of English tokens (\token) or can have one or more control blocks (\ctlblock). 
Here, the control block is the primary construct for transitions and contains \trancompplural (i.e., \startstatetag, \enstatetag, \condtag, and \acttag). 
Figure \ref{fig:workflow} illustrates an example control block with multiple condition spans (\condtag), start state span (\startstatetag), and action spans (\acttag). 
Moreover, our grammar supports nested spans and facilitates extracting complex logical relationships among them, as demonstrated in the examples in Figures \ref{fig:workflow}, \ref{fig:annotation-nested-cond}, and \ref{fig:annotation-nested-ctl}. 

\begin{figure}[t]
	\begin{center}
		\includegraphics[width=\linewidth]{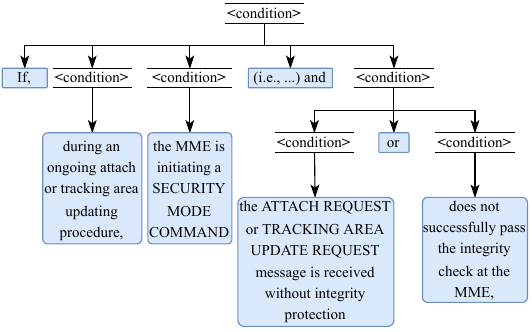}
		\caption{Nested condition tags.}
		\label{fig:annotation-nested-cond}
	\end{center}
\end{figure}

\begin{figure}[t]
	\begin{center}
		\includegraphics[width=\linewidth]{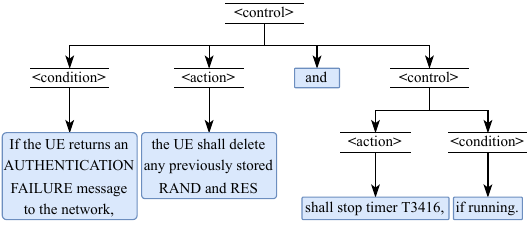}
		\caption{Nested control blocks.}
		\label{fig:annotation-nested-ctl}
	\end{center}
\end{figure}

\subsection{Constituency Parsing - \ntce}
\label{sec:ntce}

We build \ntce, a neural constituency parser that extracts \ctlblock and corresponding \trancompplural from natural language texts. As cellular specifications contain rich and complex associations of transition components, which pose a significant challenge for existing NLP models trained on general domains, we propose the following innovations to adapt and enhance state-of-the-art constituency parser~\cite{zhang2020fast} for our task:

(i) Instead of the original RoBERTa model \cite{roberta} that cannot recognize most cellular-specific technical words and phrases, we pretrain a new embedding model,  
\securoberta, from scratch with cellular-specific documents, as described in \S\ref{ref:subsec_constituency_pt}. This ensures that the constituency parser can correctly comprehend the tokens used in cellular specification documents.

(ii) We reformed the loss function so that it properly integrates span semantics and span structure information during objective optimization. We demonstrate that this is critical in improving performance in detecting the constituents.

\noindent\textbf{Workflow.}
\ntce works in a two-stage scheme, as shown in Figure \ref{fig:const_architecture}. In the \textbf{Bracketing} stage, the parser takes an input sentence and outputs an unlabeled tree housing all the constituents. In the \textbf{Labeling} stage, these constituents are classified into different labels to get the output tree. In this way, we not only get the tags of different constituents for the input, but also preserve the structural association.

\noindent\textbf{Bracketing.} Given a sentence $x$, the conditional probability of an unlabeled tree $y$ is calculated as:
    $P(y|x) = \frac{\exp^{S(x,y)}}{\sum_{y^\prime}{\exp^{S(x,y^\prime)}}} \quad$,
where $S(x,y) = \sum_{(i,j) \in y}s(i,j)$ is the sum of the scores of all constituent spans $(i,j)$ under the tree $y$, and $y^\prime$ enumerates all possible trees using the CKY algorithm~\cite{hopcroft2001introduction}.
For each word $w_i$, we use multi-layer perceptrons over domain-specific \securoberta embeddings 
to produce left boundary representations $u_l^i$ and right boundary representations $u_r^i$ for this word.
Then, for a given span $(i,j)$, we calculate the biaffine score using the left boundary word representation $w_i$ and right boundary word representation $w_j$ to determine the score for the span:
$s(i,j) = {u_l^i}^{\!\!\top} W u_r^j + b$,
where $W$ and $b$ are trainable parameters. After calculating the score of all possible trees, the best tree $\hat{y}$ is chosen as:
    $\hat{y} = \arg \max_{y} P(y|x)$.

\begin{figure}[t!]
    \centering
    \includegraphics[width=0.85\linewidth]{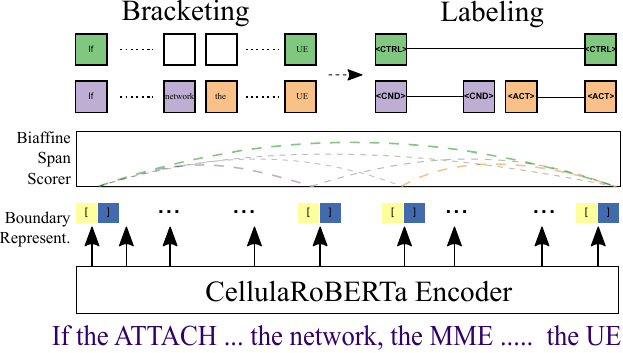}
    \caption{Detailed architecture of \ntce.}
    \label{fig:const_architecture}
\end{figure}
\noindent\textbf{Labeling.} Given an input sentence $x$ and a tree $y$, the label is predicted for each of the constituents $(i,j)$ in the tree independently. 
We denote a label as $l \in \mathcal{L}$, where $\mathcal{L}$ is the label set 
in our grammar.
For a constituent $(i,j)$ and a label $l$, we calculate a biaffine score $s(i,j,l)$ 
similarly as 
done
in the bracketing stage, and we have $|\mathcal{L}|$ biaffine heads for all the labels. We pick the best label as:
$\hat{l} = \arg \max_{l \in \mathcal{L}} s(i,j,l)$.

\noindent \textbf{Training.} 
Given a sentence $x$ with the corresponding ground truth constituency tree $y$ and label $l$, the vanilla loss \cite{zhang2020fast} is calculated as 
    $L =  L_{\text{bracket}} + L_{\text{label}}$,
where $L_{\text{bracket}}$ denotes the global conditional random field (CRF) loss 
in sentence level, and $L_{\text{label}}$ is the cross entropy loss for labeling using teacher forcing by training only with ground truth trees.
However, 
we found that $L_{\text{bracket}}$, which encodes the bracketing loss to encode the tree structure, does not work well in the early stages of training, and the model quickly converges to local minima where it only generates trivial span prediction with suboptimal performance (labeled F1 score of only 0.2849).

Consequently, we postulate that the constituent label information from $L_{\text{label}}$ could enhance the integration of span semantics that is lacking in the early training phase, which can later facilitate the model to learn structured information by training with $L_{\text{bracket}}$.
Thus, we emphasize labeling training in the early stage by modifying the loss as $L = \lambda_1 L_{\text{bracket}} + \lambda_2L_{\text{label}}$. 
In the first few stages, when bracket loss is unstable, we give high weights to $L_{\text{label}}$ (i.e., $\lambda_2 > \lambda_1$). Later, we normally train with $\lambda_2 = \lambda_1$. 
This ensures smooth training, resulting in accurate constituency parsing.

\subsection{Pretraining \securoberta}
\label{ref:subsec_constituency_pt}
The original RoBERTa~\cite{roberta} model used in  prior constituency parser \cite{zhang2020fast} fails to perform satisfactorily on cellular documents because of many unknown tokens and new domain knowledge, such as the frequent use of 3GPP-specific words, abbreviations, and technical terms. Therefore, we pretrain a RoBERTa model from scratch on cellular documents so that it can accurately generate embeddings for cellular specification tokens. 
Accordingly, we curate a large dataset of cellular documents consisting of more than 22,000 documents, including cellular specifications, change requests, technical reports, etc.
We preprocess these documents by removing page headers and footers, non-ASCII characters, tables, and figures. We then pretrain a RoBERTa model on our cleaned corpus.  
In this manner, we create \securoberta, a RoBERTa-based specialized language model providing domain-specific contextualized word representations of cellular specifications.

\section{Synthesizing IR from \trancomp}
\label{sec:ir-synthesizer}

\begin{figure*}[t!]
\begin{minipage}[t][][b]{0.38\linewidth}
    \vspace{4mm}
\centering
		\includegraphics[width=\linewidth]{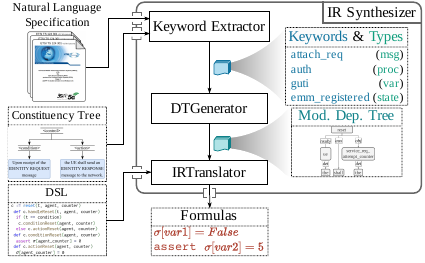}
		\caption{Architecture of \irsynthesizer.}
		\label{fig:ir-syn-architecture}
\end{minipage}
\hspace{1mm}
\begin{minipage}[t][][b]{0.60\linewidth}
\centering

		\includegraphics[width=0.98\linewidth]{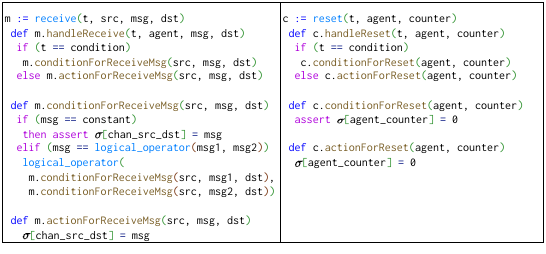}
		\caption{Definitions of DSL commands--receive and reset. 
  $\sigma$ denotes the variable store, and $\mathsf{t}$ denotes if the current use case is a condition or an action. 
  }
		\label{fig:dsl}

\end{minipage}
\end{figure*}

For each \ctlblock, \irsynthesizer takes in the transition components (\trancompplural) extracted by \ntce and generates corresponding intermediate representations (IR) as logical formulas. \fsmsynthesizer then uses the IRs to build FSMs. 

The architecture of \irsynthesizer is illustrated in Figure \ref{fig:ir-syn-architecture}. 
At first, it extracts keywords (e.g., messages, variables, events, etc.) from natural language specifications and links 
those keywords to \trancomp strings with \keyextractor. 
After that, the \astgenerator converts these strings to dependency trees, which is used by \irtranslator for mapping the \trancompplural to DSL commands and extracting the arguments in order to finally generate logical IRs (\trancompir).

\subsection{Domain-Specific Language}

For IR, we consider the conditions as 
logical expressions
(e.g., $\mathsf{chan\_ue\_mme} = \authfailure$)  
and actions as logical or arithmetic statements 
(e.g., $\mathsf{counter:=counter+1}$). 
However, these expressions and statements are often not syntactically/structurally similar to their corresponding natural language texts (e.g., ``if the UE returns \authfailure'', ``increment $\mathsf{counter}$''). As a result, direct translation from natural language to logical formulas is challenging.
Further, the lack of generalizability of existing data-driven approaches (discussed in \S\ref{sec:challenge_3})  
makes it even more challenging to
generate logical formulas directly from natural language in the cellular domain.

To address these disparities, we define a Domain Specific Language (DSL) to reliably translate \trancompplural to logical formulas.
The DSL is partially shown in Figure \ref{fig:dsl}.
\system relies on DSL because its commands can encode logical and linear integer arithmetic expressions
and logical operations, and allow the use of variables, constants, and quantifiers. 
For example, the NL string--``the \emph{UE} shall \emph{reset} the \emph{service\_req\_attempt\_counter}'' can be mapped to the DSL command $\mathsf{reset(UE, ~service\_req\_attempt\_counter)}$ and corresponding arguments, which can then be easily translated to $\mathsf{UE\_service\_req\_attempt\_counter := 0}$. 
While defining the DSL, we observe that the operations or conditional checks usually occur as action verbs 
in \trancompplural.
Thus, we extract action verbs as DSL command names 
from the specification using a neural parts-of-speech tagger~\cite{stanza},  
and 
examine the \trancompplural containing these verbs 
to identify required arguments. 
Finally, we manually define the output logical forms for each DSL rule and implement 
an interpreter to generate them.

\subsection{Keyword Linking}
Identifying arguments of DSL commands is an integral step in applying the DSL rules. 
These are various identifiers, such as variables, states, messages, etc., consisting of multiple natural language tokens, whereas an argument is a unit. 
Although 3GPP defines a few abbreviations, messages, etc., the list is not complete and often misses important identifiers (e.g., \textit{``service request attempt counter''}). 
As manual extraction of these identifiers is tedious and error-prone, 
we take an automated approach to identify
and link them in \trancomp strings with a single token keyword, which can be used as arguments to DSL commands. 
We observe that the identifiers are usually noun phrases, so we use a neural 
phrase-structure constituency parser \cite{stanza, stanza-constituency} 
to identify all noun phrases. After further filtering, processing, and applying cellular specification-specific heuristics, we extract the desired identifiers and 
assign types to them by combining automated and manual efforts.

While linking single token keywords to identifiers in \trancompplural, we observe slight variations in identifiers with the same meaning 
(e.g., ``tracking area updating accept'' and ``tracking area update accept''), 
smaller identifiers within larger ones (e.g., IMEI, IMEISV, and IMEISV request), 
typographic errors, and inconsistent uses of punctuation. 
To address these challenges, we utilize Levenshtein distance \cite{edit-distance} to correctly identify the keyword.
We also embed type information to the keywords for the next phases of \irsynthesizer.

\subsection{Constructing Dependency Trees}
\label{sec:dependency_parsing}
As discussed in \S\ref{sec:challenge_4}, DSL rules require specific arguments, and representing 
natural language transition components (\trancompplural) as a dependency tree helps identify these rules and arguments. For that purpose,
\astgenerator takes each \trancomp with clearly mapped keywords and type information as inputs. As output, it generates representative dependency trees for DSL mapping.

\noindent\textbf{Usage of dependency parse trees.} 
The DSL commands and arguments are inherently 
tree-structured~\cite{logic-tree} and can be represented with abstract syntax trees (AST). 
Therefore, generating ASTs from natural language strings is naturally the first step toward 
mapping them to DSL.
Based on this insight, \irsynthesizer first aims to map the sequential strings to DSL ASTs. 
We observe that \emph{dependency parse trees} can discover internal grammatical relations (e.g., subject-verb, verb-object, etc.) between tokens of a natural language string 
and represent it as a tree that has main verbs, e.g., ``send'', ``reset'', etc., at the central node, and the arguments at its children. 
We leverage this hierarchical structure to  
map them to ASTs as it is similar to the DSL command and argument relations in the ASTs.

To demonstrate this insight, 
we consider the string ``If the UE receives auth\_reject or tau\_reject'', whose
dependency tree is shown in Figure \ref{fig:dep_ast_trees} (a). 
Conversely, the desired DSL command is $\mathsf{receive(UE, or(auth\_reject, tau\_reject))}$,
whose AST is shown in Figure \ref{fig:dep_ast_trees} (c). 
These two trees have several common features, such as, \textit{``receive''} is the root, \textit{``UE''} is its child, and \textit{``auth\_reject''} and \textit{``tau\_reject''} are \textit{receive}'s descendants. 
Based on this observation, we utilize the dependency tree to reach DSL rules from natural language strings. 
\begin{figure}
	\begin{center}
		\includegraphics[width=0.99\linewidth]{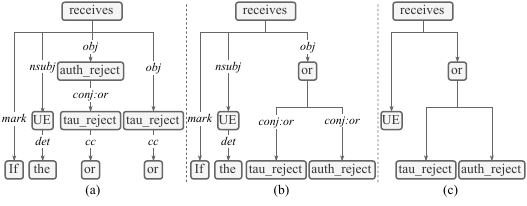}
		\caption{(a) Original dependency tree, (b) Modified dependency tree, (c) Abstract syntax tree.}
		\label{fig:dep_ast_trees}
	\end{center}
\end{figure}

\noindent\textbf{Off-the-shelf dependency parsers are not enough for cellular specifications.} 
Although dependency trees  \cite{stanford-dependency-parser, stanza} largely correspond to our translation scheme, we require some modifications before adopting them for obtaining DSL rules. 
This is because of semantic differences in how conjunctions, prepositions, or adverbs are used in typical English sentences in contrast to formal languages. 
For example, ``and'', ``or'', ``not'', etc. are generally leaves of dependency trees in natural languages, 
while these words denote logical operations 
similar to verbs in formal languages. 
To illustrate, in Figure \ref{fig:dep_ast_trees} (a), the extracted parent-child relation 
between ``auth\_reject'' and ``tau\_reject'' is not applicable in a logical translation setting.
Rather, ``auth\_reject'' and ``tau\_reject'' should be siblings under the ``or'' operation, as shown in the AST in Figure \ref{fig:dep_ast_trees} (c).

\noindent\textit{Solution.} Based on these observations, 
we perform a post-order traversal to find the logical operators 
and move the logically connected tokens under their subtree.
We show the modified dependency tree for the discussed example in Figure \ref{fig:dep_ast_trees} (b), where ``or'' is changed to the suitable position. 
Finally, the \astgenerator produces dependency trees for each \trancomp and passes them to the \irtranslator for DSL mapping.

\subsection{Synthesizing Intermediate Representation}
\label{sec:ir-translator}
The \irtranslator takes dependency trees corresponding to each transition component (\trancomp), maps these trees to DSL ASTs, and 
finally produces logical formulas or IR (\trancompir) for \trancompplural by following the  
DSL semantics.

\noindent\textbf{Mapping dependency parse tree to AST.}
A dependency tree, although similar to the target AST, does not directly represent the target DSL command. 
For that,
\irtranslator first identifies the corresponding DSL command and its arguments. 
Consequently, \irtranslator recursively traverses the dependency tree using a top-down approach, i.e., it matches the DSL command first and then the arguments. 
It also considers syntactic information from dependency trees and type information from \keyextractor  
to map the arguments based on their type and produce the correct \trancompir. 
For the example in Figure \ref{fig:dep_ast_trees}, \irtranslator first matches the rule $\mathsf{receive(t, src, msg, dst)}$ and recursively finds arguments. 
Using the subject-object relation and \textit{agent} type information, \irtranslator decides the $\mathsf{dst}$. 
The final output is 
``$\mathsf{assert}~\sigma[\mathsf{chan\_ue\_mme}] = \mathsf{auth\_reject} ~|~ \mathsf{assert}~\sigma[\mathsf{chan\_ue\_mme}] = \mathsf{tau\_reject}$'', where $\sigma$ is the variable store.

\noindent\textbf{Diverging dependency trees.}
Natural language may have many forms of text expressing the same meaning, e.g., 
``The UE, after receiving auth\_reject or tau\_reject'', 
``If auth\_reject or tau\_reject is received by the UE'', etc.
The corresponding dependency trees are structurally different as well.
The first one has ``UE'' as the left child, whereas the other one has it as the right child of ``receive''.
However, 
a reliable interpreter must generate the same \trancompir for both.
We ensure this by mapping dependency trees to DSL rules in a position-independent manner by using type information and syntactic information, e.g., part-of-speech and dependency relation. 
In this example, ``UE'' has type \textit{agent}, and ``auth\_reject'' and ``tau\_reject'' have type \textit{message}, which help map them as DSL arguments.

\noindent\textbf{Cellular specification directives.}
Cellular specifications often use the directives-- ``may'', ``should'', and ``can'', which signify that the operation is recommended or implementation-dependent~\cite{3gpp_directives}. 
Prior works \cite{doltest} identify these recommendations as ambiguities 
and potential security flaws. 
\irtranslator addresses these directives by adding uniform 
random variables to the corresponding conditions~\cite{lteinspector, 5greasoner}. 
These variables enable an automated reasoner to non-deterministically explore and test all possible options,
resulting in a comprehensive security analysis.

\noindent\textbf{Addressing missing information.}
\irtranslator works with each \trancomp,  
but it may not contain all the information necessary to generate the corresponding \trancompir. 
For instance, in Figure \ref{fig:workflow}, the condition--``the message is integrity protected'' does not mention the message name, which needs to be inferred from the overall context.
Conversely, in many cases, the \ctlblock, or even the whole paragraph, may not be sufficient for accurately inferring the information 
(e.g., actions for cause \#7 in Section 5.5.1.2.5 TS 24.301 \cite{4g_nas}).
Consequently, it needs to be inferred from previous paragraphs. 
For these reasons, we implement a context resolver in \irtranslator to find any missing information and produce the correct IR. 
For that, \irtranslator first checks the context of the concerned \ctlblock, then the whole paragraph, and follows a heuristic backtracking mechanism to search previous ones.

\noindent\textbf{Modeled actions and granularity of FSM.}
Modeled actions and the granularity of the extracted FSMs largely depend on the defined DSL rules and arguments. 
If necessary, \system can adopt more actions and more detailed modeling easily. 
In this work, we consider security-critical actions, such as sending/receiving messages, 
activating/deactivating services, incrementing counters and sequence numbers, starting timers, etc., totaling \numdsl DSL commands  
provided in \cite{hermes_artifact}. 
In contrast, limitations in our logical abstraction 
do not allow us to model some actions, e.g., list modifications (adding or removing).

Further, \system models the specifications with more details than existing works~\cite{lteinspector, 5greasoner} by considering more message fields and the corresponding protocol behavior. 
\system also models 13 services (e.g., emergency bearer, EPS, etc.), 
11 modes (e.g., EMM-CONNECTED, etc.), etc., along with states, procedures, counters, sequence numbers, timers, etc. 
However, compared to prior works~\cite{formal-5g-aka, formal-5g-auth, formal-5g-handover, formal-5g-key}, \system does not model interactions among multiple base stations or multiple network functions in the core network. 
Instead, \system abstracts these interactions because the analyzed specifications mainly involve interactions between UE and AMF (or MME) and between UE and gNB. 
Certain procedures, such as handover, span multiple documents~\cite{formal-5g-handover}. 
Although \system has the ability to extract an FSM from multiple documents,  the current instantiation of \system extracts FSMs from individual documents. As a result, 5G NAS and 5G RRC FSMs provide a partial representation of the 
handover procedure.

\section{Synthesizing FSM}
\label{sec:fsm_synthesizer}

The \fsmsynthesizer combines the intermediate representations corresponding to the transition components (\trancomp) generated by \irsynthesizer to produce complete transitions. It also compiles all the transitions and performs logical checks to ensure the consistency of generated FSMs.

\noindent\textbf{Constructing a transition by combining its components.}
As mentioned in \S\ref{sec:grammar-details}, \ctlblock is the primary construct for transitions, 
so \fsmsynthesizer combines the IRs of each \trancomp for a \ctlblock. 
While doing so, it deals with multiple logically connected condition and action spans. 
For instance, the example in Figure \ref{fig:workflow} contains two condition spans connected with ``or'', and the whole condition is connected with another condition using the ``and''.
\fsmsynthesizer first finds these logical relations from the constituency parse tree, combines them into a single condition expression, and compiles the actions for that transition. 
Moreover, \fsmsynthesizer identifies the start state and end state from corresponding IR from \irsynthesizer by considering them as conditions and actions, respectively. This is useful for logical connections (e.g., or, not, etc.) between the states.
For the running example in Figure \ref{fig:workflow}, after combining the outputs from the components, the transition with a condition, a start state, and an action is the output. 
Note that if the end state is not mentioned, 
the FSM remains in the same state.

\noindent\textbf{Compiling all transitions.}
To generate a complete formal model, \fsmsynthesizer follows the above procedure and constructs all possible transitions identified by \ntce 
in a given specification document. 
Further, as defined in \S\ref{sec:problem_statement}, these documents describe
protocol interactions among multiple participants (e.g., UE and AMF). 
Accordingly, \fsmsynthesizer partitions the list of transitions and generates separate FSMs for each participant.
Finally, \fsmsynthesizer outputs each FSM as a graph, 
where the states are represented as unique nodes, 
and the transitions are represented as directed edges 
from the start state node to the end state node.
We also label these edges with IR representations (\trancompir) of corresponding conditions and actions.

\noindent\textbf{Addressing split transitions.}
\system takes input in natural language at the paragraph level since each paragraph in the cellular specifications usually discusses a particular scenario (start state, conditions, and corresponding actions). 
However, actions in these documents are oftentimes mentioned across multiple paragraphs. 
For instance, in Section 5.5.1.2.4 of 4G NAS specifications~\cite{4g_nas}, 
when \attachreq is accepted, the network behavior is discussed in multiple paragraphs, 
leading to a divided transition.  
As a result,
formal analysis of 
the extracted FSM may not execute all of the actions, 
potentially failing to explore all possible paths.
We observe two cases of such complicated transition pairs 
where the start state is the same.
First, when the conditions in any two transitions $T_1$ and $T_2$ 
are logically equivalent  ($C_1\! \iff\! C_2\!$), 
we merge the transitions into one by combining the actions. 
Second, the conditions are not equivalent, but one implies the other, 
i.e., $\neg(C_1\! \iff\! C_2\!)\! \land\! (C_1\! \implies\! C_2\!)$  
(e.g., $C_1\! =\! x\! \land y$ and $C_2\! =\! x$). 
Here, whenever $C_1$ is $\mathsf{TRUE}$, both transitions $T_1$ and $T_2$ are valid.
Thus, we extend the list of actions in $T_1$ with the actions from $T_2$ 
and change the condition in  $T_2$ to $(C_2\! \land\! \neg C_1\!)$. 
This ensures that the formal analysis explores all possible paths in the FSM.

\noindent\textbf{Recovering missing transitions.}
\ntce sometimes fails to recognize a few action spans in natural language texts, 
leading to missing transitions. 
To recover such actions, 
we introduce a rule-based mechanism into \fsmsynthesizer. 
Specifically, within a \ctlblock, 
if we find a string with no \acttag, we scan it for action verbs and corresponding arguments covered in the action DSL commands of \irsynthesizer. 
If such a DSL command can be mapped,
we consider the string as an action \trancomp, generate the corresponding \trancompir with \irsynthesizer, and add it to the output FSM.

\section{Implementation}
We annotate cellular specification datasets with 4G-NAS Release16~\cite{4g_nas}, 5G-NAS Release17~\cite{5g_nas}, and 5G-RRC Release17~\cite{5g_rrc} documents according to the grammar shown in Figure \ref{fig:grammar}. 
These three annotated documents have $\sim$16,000 datapoints, 
and it required a total of $\sim$2800 man-hours to complete the annotations, 
which are executed by four cellular systems researchers and verified by two domain experts.

\begin{table}[t!]
\resizebox{\columnwidth}{!}{%
\begin{tabular}{@{}llr@{}}
		\toprule
		\textbf{Component} & \textbf{Libraries Used} & \textbf{Lines of Code}  \\
		\midrule
		Text extraction and cleaning & pdfplumber \cite{pdfplumber} & 380  \\
		\ntce & Huggingface \cite{huggingface}, NLTK \cite{nltk}   & 1200+ \\
        \keyextractor & NLTK \cite{nltk}, Stanza \cite{stanza}, Levenshtein \cite{edit-distance}  & 1343 \\
		\astgenerator  & NLTK \cite{nltk}, Stanza \cite{stanza} & 378  \\
		\irtranslator & & 3209 \\
		\fsmsynthesizer &  Z3 \cite{z3} & 1206 \\
		nuXmv transpiler & LTEInspector transpiler \cite{lteinspector} & 416 \\
		\uegen &  & 1000+ \\  
        \bottomrule
	\end{tabular}%
    }
	\caption{Implementation efforts for \system components.}
	\label{tab:implementation-details}
\end{table}

For \ntce, we use neural CRF-based constituency parsing \cite{zhang2020fast}. 
To train \ntce with the annotated documents, we use a learning rate of $lr = 5e^{-5}$ with a warmup of $0.1$. 
During the initial stages of training, when the bracketing loss is unstable, we use $\lambda_1 = 0.1$ and $\lambda_2 = 0.9$. 
During the rest of the training, we use $\lambda_1 = \lambda_2 = 0.5$, giving equal importance to both loss components.
All models are trained for $200$ epochs, and the best model is chosen based on a held-out dev set. 
The summary of the implementation efforts is shown in Table \ref{tab:implementation-details}.

\section{Evaluation}
\label{sec:evaluation}

To evaluate the performance of \system, we aim to answer the following research questions:
\begin{itemize}[noitemsep,topsep=0pt,leftmargin=0.4cm]
    \item \textbf{RQ1.} What is the efficacy of \ntce to automatically annotate a cellular specification document, and how it compares with existing baselines (\S\ref{sec:result-ntce})? 
    \item \textbf{RQ2.} What is the efficacy of \irsynthesizer and \fsmsynthesizer to 
    translate \trancomp to IR (\S\ref{sec:result-fsm-gold})?
    \item \textbf{RQ3.} What is the accuracy of the \system extracted FSM, 
    and how it compares with existing formal models (\S\ref{sec:result-extracted-fsm})?
    \item \textbf{RQ4.} How effective is the extracted FSM in identifying attacks and finding noncompliance (\S\ref{sec:results-attacks}, \S\ref{sec:results-unannotated}, and \S\ref{sec:results-dikeue})? 
    \item \textbf{RQ5.} How efficient is \system in extracting FSMs and performing security analysis with these FSMs (\S\ref{sec:efficiency})? 
\end{itemize}

\begin{table*}[ht]
\resizebox{\textwidth}{!}{%
\begin{tabular}{@{}lrrrrrrrrrrrrrrrrrr@{}}
\toprule
\multirow{2}{*}{} & \multicolumn{6}{c}{4G NAS $\rightarrow$ 5G NAS} & \multicolumn{6}{c}{5G NAS $\rightarrow$ 4G NAS} & \multicolumn{6}{c}{5G NAS $\rightarrow$ 5G RRC} \\ \cmidrule(l){2-7} \cmidrule(l){8-13} \cmidrule(l){14-19}  
                  & UP   & UR   & UF  & LP  & LR  & LF  & UP   & UR   & UF  & LP  & LR  & LF  & UP   & UR   & UF  & LP  & LR  & LF \\ \midrule
RFCNLP & 8.22 &  27.53    &   12.66  &  8.15   &  27.27   &  12.54  &  39.52    &  38.71   & 39.12    &  38.92   & 38.12 &  38.52    & 13.45    &   10.84  &  12.01  & 11.45 & 9.22 & 10.22 \\
\ntce    &   66.88   &  68.79    &  67.82   &  64.30   & 66.13    & 65.20   &  71.38    &    71.28 &  71.33   & 68.25    & 68.15 &  68.20     &  72.81   & 74.44    &  73.62   &  67.94 & 69.45 &  68.69\\ \bottomrule
\end{tabular}%
}
\caption{\ntce and RFCNLP entity tagger \cite{rfcnlp} results on cellular specifications. 
X $\rightarrow$ Y means we train on X and text on Y. \textbf{UP, UR, and UF} denotes unlabeled precision, recall, and F1-metrics. Similarly, \textbf{LP, LR, and LF} denotes labeled metrics.}
\label{tab:result-cellular}
\end{table*}

\begin{table*}[t!]
\begin{minipage}[t][][b]{0.58\linewidth}
\centering
\resizebox{\textwidth}{!}{%
\begin{tabular}{@{}lrrrrrrrrrrrr@{}}
\toprule
\multirow{2}{*}{} & \multicolumn{6}{c}{BGPv4, DCCP, LTP, PPTP, SCTP $\rightarrow$ TCP} & \multicolumn{6}{c}{BGPv4, LTP, PPTP, SCTP, TCP $\rightarrow$ DCCP} \\ \cmidrule(l){2-7} \cmidrule(l){8-13} 
                  & UP   & UR   & UF  & LP  & LR  & LF  & UP   & UR   & UF  & LP  & LR  & LF   \\ \midrule
RFCNLP & 60.03 & 55.05 & 57.43 & 49.92 & 45.78 & 47.76 & 38.75 & 39.00 & 38.88 & 33.80 & 34.02 & 33.91   \\
\ntce & 64.53 & 55.60 & 59.73 & 61.64 & 53.11 & 57.06 & 53.78 & 59.97 & 56.71 & 52.98 & 59.08 & 55.06 \\ \bottomrule
\end{tabular}%
}
\caption{\ntce and RFCNLP entity tagger results on the RFC dataset \cite{rfcnlp}.}
\label{tab:result-rfc}
\end{minipage}
\hspace{2mm}
\begin{minipage}[t][][b]{0.4\linewidth}
\centering
\resizebox{\textwidth}{!}{%
\begin{tabular}{@{}lcccc@{}}
\toprule
\multirow{2}{*}{} & \multicolumn{2}{c}{5G NAS} & \multicolumn{2}{c}{4G NAS} \\ \cmidrule(l){2-3} \cmidrule(l){4-5}
                  & Action     & Condition     & Action     & Condition     \\ \midrule
Synthesizer FSM ($M_{Syn}$)       & 93.86      & 94.45         & 92.23      & 92.24         \\
\system FSM ($M_{Hermes}$)     & 81.39      & 86.40         & 81.14      & 87.21         \\ \bottomrule
\end{tabular}%
}
\caption{Transition accuracy against $M_{Gold}$: $M_{Syn}$ for IR \& FSM synthesizers, $M_{Hermes}$ for \system.}

\label{tab:canonical-accuracy}
\end{minipage}
\end{table*}

\subsection{Effectiveness of \ntce}
\label{sec:result-ntce}
To answer RQ1, 
we compare \ntce with the BIO entity tagger in RFCNLP~\cite{rfcnlp} in terms of identifying constituency spans.
For this comparison, we report traditional precision, recall, and F1-score metrics \cite{sekine1997evalb}. 
We consider both unlabeled and labeled metrics--
in \textbf{unlabeled} comparison, we use these metrics for the detected spans without any consideration of the labels assigned to them. In \textbf{labeled} comparison, we also take into consideration the labels assigned.

\noindent\textbf{Results on cellular specifications.}
We compare RFCNLP \cite{rfcnlp} and \ntce using 3 different training and testing combinations of annotated documents reflecting the generalization capabilities of \ntce across layers, releases, and generations.
We first train \ntce on annotated 4G-NAS-Release16 specifications \cite{4g_nas} and evaluate on the 5G-NAS-Release17 \cite{5g_nas}.
Again, we train a separate \ntce model from scratch on 5G-NAS-Release17 specifications and evaluate on both 4G-NAS-Release16 and 5G-RRC-Release17 \cite{5g_rrc}. 
We consider each paragraph of the documents as separate datapoints during the training and the evaluation.

We show these results in Table \ref{tab:result-cellular}, 
and \ntce significantly outperforms the BIO entity tagger in RFCNLP. 
We find that \textbf{BIO} entity tagging in RFCNLP has no explicit span handling mechanism, and thus, it suffers in the challenging cellular dataset containing complex structure and several levels of nested constituents.
Moreover, RFCNLP breaks sentences into chunks using an off-the-shelf chunker, often providing erroneous results in highly technical cellular specifications.
In contrast, \ntce shows stable performance across different generations, protocols, and releases of the cellular specifications, reaching $64-74\%$ precision, recall, and F1 score. This enables our method to be used in future generations of cellular systems as well.
The lost accuracy is mainly due to the distribution shift resulting from the domain change (e.g., 4G to 5G or NAS to RRC) and the difficulty of the dataset stemming from the increasing degree of nested constituents, which complicates the semantic relationship between hierarchical spans.
Additionally, compared to 4G NAS specifications, we find that there is a slight accuracy drop in 5G NAS, which can be attributed to the increased difficulty (length and nesting).

\noindent\textbf{Results on the RFC dataset.}
To show \ntce's generalizability to other domains, we train and test it using the RFC dataset (TCP, DCCP, SCTP, PPTP, LTP, BGPv4) with the same dataset split as RFCNLP \cite{rfcnlp}. 
For comparison, we compute the same metrics on the predictions by RFCNLP \cite{rfcnlp-code}. 
Table~\ref{tab:result-rfc} presents these results and 
shows that although \ntce is primarily designed for cellular specifications, 
it outperforms RFCNLP even in the RFC dataset. 
This 
manifests \ntce's applicability to other domains.

\subsection{Effectiveness of IR and FSM Synthesizers}

\label{sec:result-fsm-gold}
To answer RQ2, we use human-annotated ground truth transition components instead of using \ntce, 
and leverage the \irsynthesizer and \fsmsynthesizer components of \system to synthesize 
transitions and generate FSMs ($M_{Syn}$). The annotations are from 4G NAS \cite{4g_nas} 
and 5G NAS \cite{5g_nas} specifications. We check $M_{Syn}$'s accuracy against the 
formal models manually constructed by prior works~\cite{lteinspector, 5greasoner}. 
These works, however, abstracted away a lot of details from the original specification. 
As a result, a head-to-head comparison with these models is not plausible. 
Thus, to evaluate $M_{Syn}$, we manually construct ground truth Gold FSMs 
($M_{Gold}$) with the same actions and 
scope as these prior works but matching the same abstraction as $M_{Syn}$. 
We use the ground truth annotations of the same paragraphs within the scope of $M_{Gold}$ 
to generate $M_{Syn}$. These include procedures covered by prior works, such as 
(de-)registration, service request, security mode control, etc.
Finally, we compute the accuracy of conditions, actions, and states in the transitions of $M_{Syn}$, considering $M_{Gold}$ as ground truth. The summary of the FSMs is provided in Table \ref{tab:explanation_fsm}, and details of computing this accuracy are in Appendix \ref{sec:eval_metric_fsm_appendix}.

The results of $M_{Syn}$ in Table \ref{tab:canonical-accuracy} show $92-94\%$ accuracy of transitions extracted by \irsynthesizer and \fsmsynthesizer as compared to the transitions manually crafted in $M_{Gold}$. 
This proves that our approach is better than 
the state-of-the-art framework on network protocols, RFCNLP \cite{rfcnlp}, which achieves $67-80\%$ accuracy.
The lost accuracy of $6-8\%$ in $M_{Syn}$ can be attributed to the wide diversity of natural language strings, 
imperfections in \keyextractor, 
and off-the-shelf dependency parsers. 
Instead of overfitting the DSL rules to these rare cases,
we strive to design \irsynthesizer generic to the cellular 
specifications across different generations, layers, and releases.

\begin{table}[ht]
\centering
\resizebox{\columnwidth}{!}{%
\begin{tabular}{@{}rlc@{}}
		\toprule
		\textbf{FSM} & \textbf{\trancomp Generated By} & \textbf{\trancompir Generated By}  \\
		\midrule
		Gold FSM ($M_{Gold}$) & Human & Human  \\
		
		Synthesizer FSM ($M_{Syn})$ & Human & \irsynthesizer \& \fsmsynthesizer  \\
		
		\system FSM ($M_{Hermes})$ & \ntce & \irsynthesizer \& \fsmsynthesizer  \\
		       
        \bottomrule
	\end{tabular}%
    }
	\caption{Transition extraction for FSMs used in evaluation.}
	\label{tab:explanation_fsm}
\end{table}

\def\arraystretch{1}
\begin{table*}[t]
	\centering
	\renewcommand{\arraystretch}{1}
	\fontsize{8}{8}\selectfont
	\begin{tabular}{| p{5.5cm}| p{5.5cm} | p{5.0cm}|} 	
		\hline
		\textbf{Annotated span} & \textbf{\ntce prediction}  & \textbf{Generated IR} \\
		\hline

        \texttt{<control>} \texttt{\errclr{<action>}} The network shall stop timer T3522 \texttt{</action>} \texttt{<condition>} upon receipt of the DEREGISTRATION ACCEPT message . \texttt{</condition>} \texttt{</control>} 
        &
        \texttt{<control>} The network shall \texttt{\errclr{<action>}} stop timer T3522 \texttt{</action>} \texttt{<condition>} upon receipt of the DEREGISTRATION ACCEPT message . \texttt{</condition>} \texttt{</control>} 
        & 
        \parbox[t]{5.0cm}{
            \texttt{condition:} chan\_ue\_amf = deregistration\_accept \\
            \texttt{actions:} timer\_t3522\_started = FALSE
        }
        \\ \hline

        \texttt{<control>} The network shall, \texttt{<condition>} on the first expiry of the timer T3460, \texttt{</condition>} \texttt{<action>} retransmit the AUTHENTICATION REQUEST message \texttt{</action>} \texttt{</control>} 
        & 
        \texttt{<control>} \texttt{\errclr{<condition>}} \texttt{\errclr{<condition>}} The network shall, \texttt{<condition>} on the first expiry of the timer T3460, \texttt{</condition>} \texttt{\errclr{</condition>}} \texttt{<action>} retransmit the AUTHENTICATION REQUEST message \texttt{</action>} \texttt{</control>} 
        & 
        \parbox[t]{5.0cm}{
            \texttt{condition:} timer\_t3460\_started \& timer\_t3460\_expired \& timer\_t3460\_expire\_counter = 1 \\
        	\texttt{actions:} chan\_amf\_ue = auth\_request
        }
        \\ \hline

        \texttt{<control>} \texttt{\errclr{<condition>}} If the SECURITY MODE COMMAND message can be accepted, \texttt{</condition>} \texttt{<action>} the UE shall take the 5G NAS security context indicated in the message into use. \texttt{</action>} \texttt{</control>} &
        \texttt{<control>} If \texttt{\errclr{<condition>}} the SECURITY MODE COMMAND message can be accepted, \texttt{</condition>} \texttt{<action>} the UE shall take the 5G NAS security context indicated in the message into use. \texttt{</action>} \texttt{</control>} & 
        \parbox[t]{5.0cm}{
            \texttt{condition:} accept\_sm\_command \\
        	\texttt{actions:} nas\_security\_context\_update = TRUE, nas\_security\_context\_valid = TRUE  
        }
        \\ \hline

        \texttt{<control>} \texttt{<condition>} The UE, when receiving the DEREGISTRATION ACCEPT message, \texttt{</condition>} \texttt{<control>} \texttt{\errclr{<action>}} stop timer T3519 \texttt{\errclr{</action>}} \texttt{<condition>} if running, \texttt{</condition>} \texttt{</control>} \texttt{</control>} 
        &
        \texttt{<control>} \texttt{<condition>} The UE, when receiving the DEREGISTRATION ACCEPT message, \texttt{</condition>} \texttt{<control>} \errclr{stop timer T3519} \texttt{<condition>} if running, \texttt{</condition>} \texttt{</control>} \texttt{</control>} 
        &
        \parbox[t]{5.0cm}{
            \texttt{condition:} chan\_amf\_ue = deregistration\_accept \& timer\_t3519\_started \\
        	\texttt{actions:} timer\_t3519\_started = FALSE  
        }
        
        \\ \hline
        
        \end{tabular}
	\caption{Examples of spans with incorrect \ntce prediction, but correct IR \cite{4g_nas, 5g_nas}.}
	\label{tab:correction-examples}
\end{table*}

\subsection{Accuracy of the Extracted FSM}
\label{sec:result-extracted-fsm}
We address RQ3 by extracting FSMs end-to-end with \system ($M_{Hermes}$) for 4G NAS \cite{4g_nas} and 5G NAS \cite{5g_nas} specifications. We compute the accuracy of conditions, actions, and states in \system FSM ($M_{Hermes}$) against the Gold FSM ($M_{Gold}$). 
As shown in Table \ref{tab:canonical-accuracy}, 
$M_{Hermes}$ has an overall accuracy of $81-87\%$, demonstrating the robustness of the approach because 
\system achieves a higher holistic accuracy than the F1-score of \ntce ($65-68\%$ in Table \ref{tab:result-cellular}).
This improvement can be attributed to cases 
where \ntce makes errors in identifying the conditions or actions span boundaries, 
but includes the action verbs and arguments in them (discussed in \S\ref{sec:ir-translator}). 
In these cases, the rest of the framework is robust enough to capture the logical meanings of the identified \trancompplural 
and correctly produce the transitions.
Moreover, \fsmsynthesizer identifies strings in \ctlblock without any tag and scans it to find action verbs and DSL arguments.
A few such cases are shown in Table~\ref{tab:correction-examples}. 
However, despite the efforts in \irsynthesizer and \fsmsynthesizer, we find that 
16 transitions
are missing in $M_{Hermes}$. We analyze such cases by consulting with the specification and find that they can be attributed to errors in \ntce and \irsynthesizer (details with examples are given in Table~\ref{tab:hidden-examples}).

\noindent\textbf{Conformance Testing.} To further test the extracted FSM's accuracy, 
we translate $M_{Hermes}$ to the nuXmv model 
following the procedure in \S\ref{sec:results-attacks} 
without instrumenting adversary capabilities into the model. 
Then, we translate 43 security-related conformance test cases 
provided by 3GPP to Linear Temporal Logic (LTL) properties and check them against the model. 
Violations of a property signify nonconformance/inaccuracies in $M_{Hermes}$ transitions. 
We found that the 5G NAS FSM passes in 33 test cases and 
4G NAS FSM passes in 32.

\subsection{Effectiveness of FSMs Extracted by \system}
\label{sec:results-attacks}

To answer RQ4,
we leverage the approach provided by prior works \cite{lteinspector, 5greasoner}, which automatically transpiles compatible FSMs to nuXmv \cite{nuxmv} model and incorporate Dolev-Yao adversary capabilities \cite{dolev-yao} (drop, modify, inject, or replay) into communication channels based on sent/received messages. 
These adversary-instrumented nuXmv models are amenable to formal model checking of security properties on the FSMs. 
Thus, to ensure compatibility with the transpiler, 
\system outputs FSMs for cellular specification documents in a Graphviz-like format with conditions and actions as propositional logic expressions, which are then translated to nuXmv and checked against security properties. 
After performing the model checking, nuXmv spits out traces of counterexamples when any of the properties are violated.

\noindent\textbf{Counterexample verification and vulnerability detection.}
A counterexample from nuXmv is basically a trace of message exchanges and internal variable updates leading to the violation of the property.
However, FSMs, even manually crafted ones, often include erroneous transitions \cite{lteinspector}, as is the case with \system-extracted FSMs.
Thus, after a counterexample is extracted,
we manually verify it by consulting the specification document.
Further, when we find an erroneous counterexample,
we manually refine the model and correct the related incorrect transitions.
This process is counterexample-guided model refinement, 
which facilitates further checking security properties on the refined model
and identifying more counterexamples. 
Finally, we follow prior works \cite{lteinspector,5greasoner} and 
manually model the remaining counterexamples in ProVerif \cite{proverif} to verify 
their feasibility under cryptographic assumptions.

\noindent\textbf{Experiment setup, properties, and procedures.}
We extract separate FSMs from 4G-NAS-Release16 \cite{4g_nas}, 5G-NAS-Release17 \cite{5g_nas}, and 5G-RRC-Release17 \cite{5g_rrc} specifications, 
and generate separate adversary-instrumented nuXmv models.
While extracting the 5G-NAS-Release17 FSM, we use \ntce trained with 4G-NAS-Release16 specification, and while extracting the FSMs from 4G-NAS-Release16 and 5G-RRC-Release17 specifications, we use \ntce trained with 5G-NAS-Release17 specification (the same setting as \S\ref{sec:result-ntce}).

\def\arraystretch{1}
\begin{table*}[t]
	\centering
	\renewcommand{\arraystretch}{1}
	\fontsize{8}{8}\selectfont
	\begin{tabular}{| p{3.5cm} | p{8.5cm} | p{3.5cm} |} 	
		\hline
		\textbf{Cause} & \textbf{Example} & \textbf{Number of missing transitions} \\
		\hline

        \ntce missed \texttt{<action>} &
        \texttt{<control>} \texttt{<action>} The MME shall initiate the GUTI reallocation procedure by sending a GUTI REALLOCATION COMMAND message to the UE \texttt{</action>} and \errclr{starting the timer T3450.} \texttt{</control>} & 6
        \\ \hline 
        
        \ntce merged two separate \texttt{<action>}s together &
        \texttt{<control>} \errclr{\texttt{<action>}} The MME initiates the NAS security mode control procedure by sending a SECURITY MODE COMMAND message to the UE \errclr{and} starting timer T3460 (see example in figure 5.4.3.2.1). \errclr{\texttt{</action>}} \texttt{</control>} & 2
        \\ \hline    
        
        \ntce divided unit \texttt{<action>} &
        \texttt{<control>} \texttt{<action>} the UE shall include the Uplink data status IE in the SERVICE REQUEST message, \errclr{\texttt{</action>}, or in the REGISTRATION REQUEST message. } \texttt{</control>} & 2
        \\ \hline    
        
        \ntce classified \texttt{<action>} as \texttt{condition} &
        \texttt{<control>} \errclr{\texttt{<condition>}} b) proceed with the pending procedure; and \errclr{\texttt{</condition>}} \texttt{</control>} & 2
        \\ \hline     
        
        \ntce missed \texttt{<control>} &
        The detach procedure is initiated by the UE by sending a DETACH REQUEST message. The Detach type IE included in the message indicates whether detach is due to a "switch off" or not. & 2
        \\ \hline       
        
        \astgenerator produces wrong part-of-speech tag &
        a) \errclr{enter} 5GMM-IDLE mode;  & 1
        \\ \hline    
        
        Incomplete set of DSL rules in \irsynthesizer &
        a) the UE shall check the authenticity of the core network \errclr{by means of} the AUTN parameter received in the AUTHENTICATION REQUEST message. & 1
        \\ \hline

        \end{tabular}
	\caption{Examples of missing transitions \cite{4g_nas, 5g_nas}.}
	\label{tab:hidden-examples}
\end{table*}

In this experiment, we check the 20 security properties from prior works \cite{lteinspector, 5greasoner} and 7 new properties that cover the scope of other related works \cite{practical-shaik, ltefuzz, defeating-imsi, white-rabbit, dos-yu, borgaonkar2019new, chlosta20215g, misconfiguration-commercial} written in Linear Temporal Logic (LTL). 
Moreover, for model checking, 
adversary-instrumented nuXmv models are created from the FSMs extracted from 4G-NAS-Release16~\cite{4g_nas}, 5G-NAS-Release17 \cite{5g_nas}, and 5G-RRC-Release17 \cite{5g_rrc} specifications.
These models cover 9 NAS and 10 RRC layer procedures, including registration, deregistration, configuration update, service request, RRC Setup, Reconfiguration, paging, and Resume. 
The properties and procedures are 
listed in Appendix \ref{sec:properties_procedures_appendix}.

\noindent\textbf{Results.}
The formal analysis identifies \prevattacks previous and \newattacks new vulnerabilities in the extracted FSMs.
Table \ref{tab:vulnerabilities} shows the list of attacks and summarizes the findings of the security analysis 
on the extracted FSMs.
We also list the previous vulnerabilities (within the scope of the analyzed specification) not detected by the analysis with \system FSMs. 

Among the not identified ones, attacks \#15 and \#30 are due to errors in transition component extraction by \ntce, and attack \#10 is due to underspecification of the specifications. 
For the rest of the non-identified attacks, 
a manual inspection uncovered that 
the corresponding vulnerabilities are not present in the analyzed release of the specifications, and thus,
\system does not model those behaviors. 
For example, the 5G-RRC specification was updated due to the report of the AKA-Bypass attack \cite{ltefuzz}, so the version we analyzed does not contain the related vulnerability.
In addition, due to the imperfection in the extracted FSM, 
the counterexample verification process, 
as discussed above, 
finds \attacksfp erroneous counterexamples from nuXmv model checker. 
We correct 47 transitions to refine the model through the counterexample-guided refinement process.
On the other hand, 
attacks \#20-21 and \#35 are new attacks identified using \system FSMs. 
These new attacks are identified using properties from prior works \cite{lteinspector, 5greasoner},
which further bolsters that \system-generated FSMs exhibit larger scopes and details than prior manually crafted models. 
Note that similar issues as 20 and 21 are concurrently reported in change requests\cite{cr_nas_count, cr_cag_list}, but these reports do not discuss the exploits.

\begin{table}[t!]
\resizebox{\columnwidth}{!}{%
\begin{tabular}{@{}rlccc@{}}
            \toprule
            \textbf{ID}  &  \textbf{Attack}  & \textbf{Gen/Layer} &  \textbf{N}  &  \textbf{D}  \\ \midrule
            1  &  Downgrade to non-LTE network services \cite{practical-shaik}  & 4G NAS &  \cross  &  \tick \\
            2  &  Denying all network services \cite{practical-shaik}  & 4G NAS &  \cross  &  \tick \\
            3  &  Denying selected service \cite{practical-shaik}  & 4G NAS &  \cross  &  \cross \\
            4  &  Signaling DoS \cite{lee2009detection, bassil2013effects, leong2014unveiling, kambourakis2011attacks}  & 4G NAS &  \cross  &  \tick \\
            5  &  S-TMSI catching \cite{ltefuzz}  & 4G NAS &  \cross  &  \tick \\
            6  &  IMSI catching  \cite{defeating-imsi}  & 4G NAS &  \cross  &  \tick \\
            7  &  EMM Information \cite{white-rabbit}  & 4G NAS &  \cross  &  \tick \\
            8  &  Impersonation attack \cite{misconfiguration-commercial}  & 4G NAS &  \cross  &  \cross \\
            9  &  Synchronization Failure attack \cite{dos-yu}  & 4G NAS &  \cross  &  \cross \\
            10  &  Malformed Identity Request \cite{michau2016not} & 4G NAS &  \cross  &  \cross \\
            11  & Neutralizing TMSI refreshment \cite{5greasoner}  & 5G NAS &  \cross  &  \cross \\
            12  &  NAS Counter Reset \cite{5greasoner}  & 5G NAS &  \cross  &  \tick \\
            13  &  Uplink NAS Counter Desynchronization \cite{5greasoner}  & 5G NAS &  \cross  &  \tick \\
            14  &  Exposing NAS Sequence Number \cite{5greasoner}  & 5G NAS &  \cross  &  \tick \\
            15  &  Cutting off the Device \cite{5greasoner}  & 5G NAS &  \cross  &  \cross \\
            16  &  Exposure of SQN \cite{borgaonkar2019new}  & 5G NAS &  \cross  &  \tick \\
            17  &  5G AKA DoS Attack \cite{survey-cao}  & 5G NAS &  \cross  &  \tick \\
            18  &  SUCI catching \cite{chlosta20215g}  & 5G NAS &  \cross  &  \cross \\
            19  &  IMSI cracking \cite{torpedo}  & 5G NAS &  \cross  &  \cross \\
            20  &  NAS COUNT update attack  & 5G NAS &  \tick  &  \tick \\
            21  &  Deletion of allowed CAG list  & 5G NAS &  \tick  &  \tick \\
            22  &  Downgrade using \attachreject / \regreject \cite{practical-shaik}  & 4G/5G NAS &  \cross  &  \tick \\
            23  &  \authreject attack \cite{dos-yu}  & 4G/5G NAS &  \cross  &  \tick \\
            24  &  \detachreq / \deregreq attack \cite{lteinspector}  & 4G/5G NAS &  \cross  &  \tick \\
            25  &  \servicereject attack \cite{practical-shaik}  & 4G/5G NAS &  \cross  &  \tick \\
            26  &  Denial-of-Service with \rrcsetupreq attack \cite{5greasoner}  & 5G RRC &  \cross  &  \cross \\
            27  &  Installing Null Cipher and Null Integrity \cite{5greasoner}  & 5G RRC &  \cross  &  \tick \\
            28  &  Lullaby Attack \cite{5greasoner}  & 5G RRC &  \cross  &  \tick \\
            29  &  Incarceration with \rrcreject and \rrcrelease \cite{5greasoner}  & 5G RRC &  \cross  &  \tick \\
            30  &  Measurement report \cite{practical-shaik}  & 5G RRC &  \cross  &  \cross \\
            31  &  RLF report \cite{practical-shaik}  & 5G RRC &  \cross  &  \tick \\
            32  &  Blind DoS attack \cite{ltefuzz}  & 5G RRC &  \cross  &  \cross \\
            33  &  AKA bypass\cite{ltefuzz}  & 5G RRC &  \cross  &  \cross \\
            34  &  Paging channel hijacking \cite{lteinspector}  & 5G RRC &  \cross  &  \cross \\
            35  &  Energy Depletion with \rrcsetup  & 5G RRC &  \tick  &  \tick \\ \bottomrule
        \end{tabular}%
            }
	\caption{Vulnerabilities identified in \system extracted FSMs. N: New, D: Detected. 
 }
	\label{tab:vulnerabilities}
\end{table}

\subsubsection{Identified Three New Vulnerabilities}
\label{sec:new-attacks}

\noindent\textbf{(1) Deletion of allowed CAG list.}
When a UE in a closed access group (CAG) cell 
sends a \regreq message through a gNB, 
the AMF determines whether any of the CAG ID(s) from the gNB 
are in the list of Allowed CAGs for that UE. 
In case this check fails,
the AMF rejects the registration and sends a \regreject message with 5GMM cause \#76.
Upon receiving this message, if the integrity check is successful, the UE deletes its allowed CAG list \cite{5g_nas}.
An adversary can exploit this behavior and force the UE to delete the allowed CAG list, resulting in a denial of service (DoS).

\begin{figure}[h]
	\vspace{2.5mm}
	\begin{center}
		\includegraphics[width=0.85\linewidth]{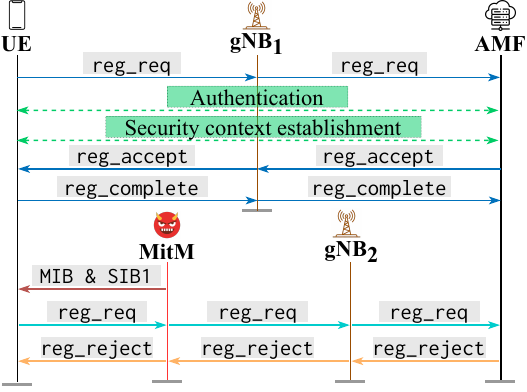}
		\caption{Deletion of allowed CAG list.}
		\label{fig:attack_cag}
	\end{center}
	\vspace{2.5mm}
\end{figure}

\noindent\textit{\underline{Root cause.}} 
The AMF sends the \regreject based on the received information from the gNB. 
However, the UE has no means to verify the CAG IDs for which the AMF sends the \regreject message. 
Without this knowledge, the UE cannot identify if there is a mismatch in CAG ID(s) when it receives the \regreject with 5GMM cause \#76.
Thus, the vulnerability stems from a design flaw.

\noindent\textit{\underline{Attack.}} 
The adversary, at first, lets the victim UE connect to a benign base station (gNB$_1$) and complete registration with the AMF. 
After that, 
the adversary 
lures the victim to connect to a machine-in-the-middle (MitM) fake base station. 
She then finds a gNB with a different set of CAG IDs (gNB$_2$) and relays the \regreq to it (Figure \ref{fig:attack_cag}). 
gNB$_2$ forwards the message to the benign AMF, and the AMF checks the CAG list from gNB$_2$.
After finding a mismatch, 
the AMF responds with integrity-protected \regreject with 5GMM cause \#76.
By forwarding this message to the UE, the adversary forces it to delete the allowed CAG list. 
Note that the initial registration through gNB$_1$ ensures that the AMF responds with integrity-protected \regreject. Without integrity protection, the UE does not accept this message \cite{5g_nas}.

\noindent\textit{\underline{Impact.}} 
The deletion of the allowed CAG list results in DoS.
Without the allowed CAG list, 
the UE cannot get access to the non-public network (NPN) that it is originally permitted to. 
Repeating this attack will cause a prolonged DoS.

\noindent\textbf{(2) Energy Depletion using \rrcsetup.}
The RRC resume procedure, introduced in 5G, strives to reduce power consumption and connection setup overhead. 
However, during this procedure, the UE accepts insecure \rrcsetup messages in response to \rrcresumereq. 
Upon receiving this message, the UE deletes RRC security contexts and releases radio bearers. 
Reestablishing a secure RRC connection requires expensive cryptographic operations, leading to energy depletion.

\noindent\textit{\underline{Root cause.}} 
The root cause of this vulnerability can be attributed to the specification allowing non-integrity-protected \rrcsetup messages in response to the \rrcresumereq \cite{5g_rrc}. 

\begin{figure}[t]
	\begin{center}
		\includegraphics[width=0.8\linewidth]{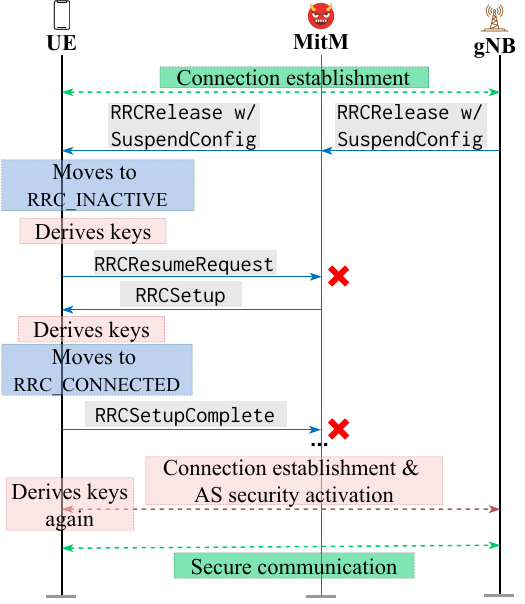}
		\caption{Energy depletion using RRCSetup.}
		\label{fig:attack_rrcsetup}
	\end{center}
\end{figure}

\noindent\textit{\underline{Attack.}} 
As the \rrcsetup messages are not integrity protected nor ciphered, 
an adversary can easily craft a plaintext \rrcsetup message.
Thus, she listens for an \rrcresumereq message in the communication channel.
After that, as shown in Figure \ref{fig:attack_rrcsetup}, when she detects an \rrcresume message from the legitimate base station,
she drops the message and spoofs a plaintext \rrcsetup. 
As a result, 
upon receiving this message,
the UE deletes its RRC security contexts and releases radio bearers.

\noindent\textit{\underline{Impact.}} 
By interrupting the RRC resume procedure, 
the adversary forces the UE to perform redundant, expensive cryptographic operations to set up the security context again. 
This leads to extraneous energy consumption. 
An adversary can repeatedly perform this attack and cause battery depletion.

\noindent\textbf{(3) NAS COUNT update attack.}
5G NAS specification \cite{5g_nas} allows a UE to store the estimated NAS COUNT, i.e., 
the NAS COUNT used in the \underline{\emph{last}} successfully integrity-verified NAS message. 
However, the adversary can exploit this behavior to manipulate the NAS COUNT and perform replay attacks.

\noindent\textit{\underline{Root cause.}} 
The root cause is a design flaw-- UE stores the NAS COUNT of the \underline{\emph{last}} integrity-verified message \cite{5g_nas}.
This allows the UE to accept integrity-protected NAS messages that have been delayed and are received out of sequence.

\noindent\textit{\underline{Attack.}} 
The adversary sets up a MitM fake base station (gNB) between the victim UE and a benign  AMF. 
Then, she can arbitrarily delay any message to update the NAS COUNT to a lower value.
For example, in Figure \ref{fig:nas_count}, 
the NAS COUNT of msg$_1$ is $X||a$, where $X$ is the overflow counter, $a$ is the sequence number, and $||$ denotes concatenation.
The adversary delays msg$_2$ (NAS COUNT: $X||a+1$)
and sends it after msg$_4$ (NAS COUNT: $X||a+3$).
This message (msg$_2$) is valid and integrity-protected by the AMF.
Thus, this out-of-sequence message
causes the victim to update its NAS COUNT to $X||a+1$. 
This opens the scope of 
replayed messages with NAS COUNTs $X||a+2$ and $X||a+3$.

\noindent\textit{\underline{Impact.}} 
As the UE becomes vulnerable to replayed NAS messages, 
the adversary can exploit this to perform traceability attacks to extract a victim's presence in a target area \cite{traceability-1, formal-5g-auth, 5greasoner}. 
This violates security requirements in 5G networks \cite{5g_sec}.

\subsection{Analysis of Unannotated Document}
\label{sec:results-unannotated}
We also analyze an unannotated document, i.e., 
5G-NAS-Release15~\cite{5g_nas_rel15}, by using  
\ntce trained on the 4G-NAS-Release16 and extracting the FSM of 5G-NAS-Release15. 
Following the formal analysis technique discussed in \S\ref{sec:results-attacks}, 
we check the same properties and procedures. 
This analysis finds the attacks migrating from 4G to 5G (attacks \#22-25 in Table \ref{tab:vulnerabilities}), and five previous 5G attacks (attacks \#12-14 and \#16-17 in Table \ref{tab:vulnerabilities}). 
The analysis also found 5 spurious counterexamples, handled similarly as discussed in \S\ref{sec:results-attacks}.

\begin{figure}[]
	\begin{center}
		\includegraphics[width=0.8\linewidth]{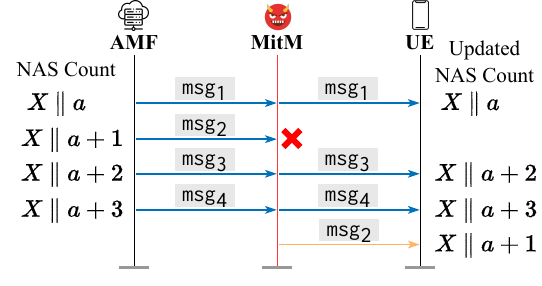}
		\caption{NAS COUNT update attack.}
		\label{fig:nas_count}
	\end{center}
\end{figure}

\begin{table}[t!]
\resizebox{\columnwidth}{!}{%
\begin{tabular}{@{}rlc@{}}
		\toprule
		\textbf{ID} & \textbf{Deviation} & \textbf{Detected}  \\
		\midrule
		1 & Replayed GUTI reallocation at specific sequence & \tick  \\
		
		2 & Replayed GUTI reallocation anytime & \tick  \\
		
		3 & Plaintext \authreq & \cross  \\
		
		4 & Plaintext \identityreq & \cross  \\
		
		5 & Selective replay of \smcommand & \tick  \\
		
		6 & GUTI reallocation before attach procedure completed & \tick  \\
		
		7 & \authresponse after \smreject & \tick  \\
		
		8 & \authfailure reply & \tick  \\
		
		9 & Replayed \smcommand & \tick  \\

        \bottomrule
	\end{tabular}%
    }
	\caption{Identified deviations in 4G UE implementations \cite{dikeue}. 
 }
	\label{tab:results_dikeue}
\end{table}

\subsection{Analysis of Cellular Implementations}
\label{sec:results-dikeue}

To further address RQ4,
we analyze the security of 4G NAS FSM \cite{4g_nas} of \numdevices cellular baseband implementations ($M_I$) 
using $M_{Hermes}$. 
We run DIKEUE's equivalence checker~\cite{dikeue} between 
$M_{Hermes}$ and $M_I$ and detect exploitable deviations in $M_I$ from $M_{Hermes}$. 
As shown in Table~\ref{tab:results_dikeue}, $M_{Hermes}$ identifies \numdeviations deviations, which 
were also identified by DIKEUE.  
However, the other two deviations that $M_{Hermes}$ misses (\#3 \& \#4), but DIKEUE detects result from 
conflicts between two separate documents~\cite{dikeue}.  
In addition, due to imperfect FSM, we face \deviationsfp 
false positive deviations, which we manually verify.

\subsection{Efficiency of \system}
\label{sec:efficiency}
To address RQ5, we compute the time required for FSM extraction and subsequent security analysis by \system. 
On an NVIDIA RTX A6000 machine, training \ntce takes $\sim$4 hours, and prediction takes less than a minute for each document.
\keyextractor
takes 8-12 hours of computation on a machine with an Intel Core i7-10610U processor and 48 GB RAM and around 5 hours of manual effort. 
\irsynthesizer and \fsmsynthesizer require 1.5-2 hours, and formal model checking requires 1-2 minutes. 
Finally, manually verifying the counterexamples requires $\sim$5 hours for each FSM. These requirements are detailed in Table \ref{tab:time}.

\begin{table}[t!]
\resizebox{\columnwidth}{!}{%
\begin{tabular}{@{}llr@{}}
    \toprule
    \textbf{Component} & \textbf{Document} & \textbf{Time required}  \\
    \midrule
    
    \multirow{2}{4 cm}{\ntce training} 
    & 4G-NAS-Release16 & 4 hour 3 minutes  \\
    & 5G-NAS-Release17 & 4 hour 9 minutes  \\
    \midrule

    \multirow{4}{4 cm}{\ntce prediction} 
    & 4G-NAS-Release16 & 42 seconds  \\
    & 5G-NAS-Release17 & 33 seconds  \\
    & 5G-NAS-Release15 & 16 seconds  \\
    & 5G-RRC-Release17 & 19 seconds  \\
    \midrule
    
    \multirow{3}{4 cm}{\keyextractor computation} 
    & 4G-NAS-Release16 & 13 hour 6 minutes  \\
    & 5G-NAS-Release17 & 13 hour 39 minutes  \\
    & 5G-RRC-Release17 & 7 hour 8 minutes  \\
    \midrule    
    
    \multirow{3}{4 cm}{Keyword type assignment} 
    & 4G-NAS-Release16 & 5 hour 27 minutes  \\
    & 5G-NAS-Release17 & 5 hour 19 minutes  \\
    & 5G-RRC-Release17 & 5 hour 46 minutes  \\
    \midrule  
    
    \multirow{4}{4 cm}{\irsynthesizer \& \fsmsynthesizer} 
    & 4G-NAS-Release16 & 43 minutes  \\
    & 5G-NAS-Release17 & 43 minutes  \\
    & 5G-NAS-Release15 & 35 minutes  \\
    & 5G-RRC-Release17 & 12 minutes  \\
    \midrule

    \multirow{4}{4 cm}{Formal model checking} 
    & 4G-NAS-Release16 & 78 seconds  \\
    & 5G-NAS-Release17 & 117 seconds  \\
    & 5G-NAS-Release15 & 106 seconds  \\
    & 5G-RRC-Release17 & 67 seconds  \\
    \midrule
    
    \multirow{4}{4 cm}{Counterexample verification} 
    & 4G-NAS-Release16 & 5 hour 24 minutes  \\
    & 5G-NAS-Release17 & 5 hour 39 minutes  \\
    & 5G-NAS-Release15 & 6 hour 16 minutes  \\
    & 5G-RRC-Release17 & 2 hour 34 minutes  \\
    \bottomrule
\end{tabular}%
}
\caption{Time requirement for FSM extraction and subsequent analysis.}
\label{tab:time}
\end{table}

\section{Related Work}

\noindent\textbf{Security Protocol Analysis Using NLP Methods.} 
RFCNLP \cite{rfcnlp} extracts FSMs from RFC documents with a BIO entity tagging mechanism. However, it fails to address nested constituents nor identify control blocks. 
Some works \cite{protocol-disambiguation, spec-fuzzing} also parse RFCs to extract source code, ambiguities, or test cases.
Chen et al. \cite{nlp-payment} analyze payment system guidelines to find logical flaws.
Several works also leverage explicit rules to extract permissions, inputs, or ontology from software documentation \cite{whyper, dase, text-mining-software}. 
Atomic \cite{bookworm-game} finds proof-of-concept of vulnerabilities using hazard identifiers, 
while CREEK \cite{sr-cr} identifies security-relevant change request documents.
These works do not extract any formal model of the analyzed system.
Another group of works uses NLP tools and program analysis on source code to extract logical errors or formal models \cite{auto-reverse, c2s, zhang-ltl}.
In contrast, \system aims to build a formal representation from the specifications.

\noindent\textbf{Text-to-logic conversion Using NLP.} 
Several works use semantic parsing to transform natural language into logical forms, such as lambda calculus \cite{learning-zettlemoyer}, meaning representation \cite{banarescu2013abstract}, programming language \cite{zhong2017seq2sql, spider, learning-iyer}, and first-order logic \cite{han2022folio}. 
Variations of neural encoder-decoder models have been adopted for these tasks\cite{language2logic-attn, exploring-logic, structvae}. 
However, these works are not easily applicable to the extraction of formal representations from cellular specifications 
because of the domain gap, lack of cellular text-to-logic dataset, or increased complexity. 

\noindent\textbf{Security Analysis of Cellular Specifications.} 
Several works manually build formal models from cellular specifications \cite{formal-5g-aka, formal-5g-auth, formal-5g-key, formal-5g-handover, 5greasoner, lteinspector}, 
and use model checkers to detect vulnerabilities. 
Some works also extract formal models from white-box \cite{prochecker} or black-box \cite{dikeue} cellular implementations. 
In contrast, \system generates a formal model from the specifications.
Further, LTEFuzz \cite{ltefuzz} generates test cases from security properties. 
DoLTEst \cite{doltest} provides a comprehensive negative testing framework for UE implementations using properties and security-relevant states.
Moreover, several works test baseband implementations \cite{basespec, basesafe, baseband-1, baseband-2}.
These works focus on testing cellular systems but do not build formal models, which is the primary objective of this work. 

\section{Discussion and Limitations}

\noindent\ding{111} \textbf{Underspecifications, ambiguities, and conflicts.}
\system does not deal with identifying underspecifications (e.g., implicit reject, unspecified reject cause, etc.~\cite{doltest}), 
and we consider this orthogonal to this work.
For ambiguities stemming from directives, e.g., ``should'', ``may'', etc.,  in 
cellular specifications, 
\system utilizes uniform random variables, 
and resulting security analysis can consider all possibilities 
equally likely, uncovering potential security flaws.
For ambiguities or inconsistencies originating from table information or information from multiple specification documents, we consider these as our future work as we currently do not incorporate table data into the generated FSM or perform security analysis on FSMs from multiple documents.

\noindent\ding{111} \textbf{Multiple documents.}
One of the challenges of extracting FSMs for cellular systems is that the protocols are described across multiple documents. Although \system has the capabilities to handle text from multiple documents and extract the corresponding transitions constituting the full FSM, we restrict our experiments to single documents to facilitate comparison with prior works \cite{rfcnlp}. In the future, we plan to incorporate multiple documents and evaluate \system.  

\noindent\ding{111} \textbf{Manual efforts.}
In \keyextractor, 
after automatically extracting the keywords from a document and assigning types to most of them,  
the user needs to assign types for 
the rest. 
This effort is significantly less than manually extracting and typing \emph{all} the keywords from scratch. 
For new functionalities or operations, the user needs to define 
the syntax and semantics for 
new DSL rules. 
Furthermore, security analysis on the extracted models requires manual efforts to define security properties and to analyze model checker outputs. 
In case a spurious counterexample is identified, relevant transitions are manually traced and fixed according to the specifications.
Further, to check the feasibility of the counterexamples, we manually model them in ProVerif~\cite{proverif}.

\noindent\ding{111} \textbf{Utility of \system generated imperfect FSM.}
In the absence of 
3GPP-defined standard formal models,
prior research focuses on 
manually modeling the specifications and verifying security properties~\cite{formal-5g-auth, formal-5g-aka, formal-5g-handover, formal-5g-key, lteinspector, 5greasoner}. 
However, due to frequent updates, changing the formal models demands strenuous efforts.
Conversely, although imperfect,
\system-generated FSMs can 
significantly reduce this effort.
When the formal model checker provides any output, we manually verify them by consulting the specifications and detecting any errors. 
In case of errors, 
we track the relevant transitions 
in the FSM 
and correct them according
to the specification. 
Compared to manually building the FSM from scratch, which may also entail errors \cite{lteinspector}, 
this effort is reasonable. 

\noindent\ding{111} \textbf{No gold FSM.} 
If the standard body, i.e., 3GPP provides a gold FSM and it is sufficiently detailed for comprehensive formal analysis, 
it could potentially provide better results than hand-crafted or automatically extracted FSMs. 
However, 3GPP does not provide any such FSM for cellular protocols. 
In fact, 3GPP plans against that 
since constructing such a gold FSM is extremely tedious, often error-prone,  
and may cause inconvenience in handling interoperabilities among stakeholders as the 
specifications leave many details to the developers' discretion.

\section{Conclusion and Future Work}
We present \system, a novel framework that automatically extracts FSMs from cellular specifications. 
Its \ntce, leveraging \securoberta, provides large improvements to capture transition components in natural language compared to existing works. The transitions generated by \system achieve 81-87\% accuracy, and security analysis on the extracted FSMs enables us to identify \newattacks new attacks and \prevattacks previous attacks exploiting design flaws. For future work, we will extend \system to detect underspecifications and conflicting sentences in cellular specifications. We will also incorporate multiple documents while extracting FSMs with \system.

\section*{Acknowledgements}
We thank the anonymous reviewers and the shepherd for their feedback and suggestions. We also thank GSMA for cooperating with us during the responsible disclosure. 
This work has been supported by the NSF under grants 2145631, 2215017, and 2226447, the Defense Advanced Research Projects Agency (DARPA) under contract number D22AP00148, and the NSF and Office of the Under Secretary of Defense-- Research and Engineering, ITE 2326898, as part of the NSF Convergence Accelerator Track G: Securely Operating Through 5G Infrastructure Program.

\bibliographystyle{plain}
\bibliography{rfc-analysis}

\appendix

\section{Explanation of Evaluation Metric To Compare FSMs} \label{sec:eval_metric_fsm_appendix}
Given a paragraph of the specification, we would have a set of ground truth transitions $\mathsf{S_{GT}}$ and a set of inference transitions $\mathsf{S_{I}}$. Now, the Boolean expression present as the condition in these transitions can be complex, and inferring quantitative accuracy from it would be difficult. To take the conditions into a consistent form, we first convert the conditions of all transitions in both $\mathsf{S_{GT}}$ and $\mathsf{S_{I}}$ into disjunctive normal form (DNF). 

After this conversion, we observe that if a condition has two or more terms joined with disjunction operators, satisfying any one of the terms suffices. As such, if an inference transition has only one of the terms present as the condition, also with the correct set of actions, it may be wrongly penalized. Thus, we split such transitions of both sets into multiple transitions with the same actions. The obtained transitions will have the individual product terms of the DNF obtained as their condition. 

Next, we observe that many of the specification lines are applicable irrespective of the present state, while some particular lines correspond to transitions going from a certain start state or reaching a particular end state. Thus, to capture the state information in transitions where certain start and end states are present, we append the matching of the start state to the transition's condition and consider the reaching of an end state as an action.

After these conversions, we have a revised set of transitions $\mathsf{S'_{GT}}$ and $\mathsf{S'_{I}}$, which are equivalent to $\mathsf{S_{GT}}$ and $\mathsf{S_{I}}$, respectively. For each of the ground truth transition $ T_{GT} \in \mathsf{S'_{GT}} $, we compare it with each inferred transition $ T_{I} \in \mathsf{S'_{I}} $. We obtain a match score for a transition set $\mathsf{S'_{I}}$ in two ways:

\begin{itemize}[noitemsep,topsep=0pt,leftmargin=0.4cm]
    \item \textbf{Conditions.} We look at the condition of a ground truth transition $T_{GT} \in \mathsf{S'_{GT}}$, and compare its condition expression with the condition of each inferred transition $ T_{I} \in \mathsf{S'_{I}}$. Note that the conditions are now terms combined with conjunction operators. We count the number of terms in the ground truth condition $c_{GT}$ of $T_{GT}$ that is accurately present (i.e., present and also the value is correct) in the inference transition $T_{I}$'s condition $c_{I}$. We divide this count with the number of terms present in $c_{GT}$ to obtain a \textit{match score} for each of the inference transitions $T_{I}$. Thus, the match score is
    $$\frac{|m_{GT, I}|}{|c_{GT}|}$$
    where $|m_{GT, I}|$ is the number of terms that are both present in the ground truth and the inferred transition's condition with the same value, and $|c_{GT}|$ is the number of terms in $T_{GT}$'s condition. 
 
    As an example, consider a ground truth transition $T_{GT}$ has the condition $a\And b\And c = D$ and an inferred transition $T_{I}$ has the condition $a \And c=E \And !d$, where $a,~b,~d$ are boolean variables, and $c$ is an enumeration variable. Here, the inferred transition has the terms $a$ and $c$ present, but $c$ is present with a different value than the ground truth. Here, the match score is $\frac{1}{3}$.
	
    We take the highest match score among all inferred transition $ T_{I} \in \mathsf{S'_{I}} $, and consider it as the score corresponding to the ground truth transition $T_{GT}$. We report the average score obtained for all ground truth transitions as the FSM accuracy for the set of transitions in $\mathsf{S_{I}}$ in terms of conditions.
	
    \item \textbf{Actions.} For actions, we consider each action $a_{GT} \in T_{GT}$ for all ground truth transitions $T_{GT} \in \mathsf{S'_{GT}} $. Next, we only consider the inference transitions $T_{I}$ where $a_{GT} \in T_{I}$. We again compare the condition expressions for all the obtained $T_{I}$, and report the best scoring one as the match score corresponding to $a_{GT}$. We report the average score for all such actions $a_{GT}$ as the accuracy of $S_I$ in terms of actions.
	
    As an example, suppose we have a ground truth transition set $S'_{GT} = \{ T_1 \}$ and the inference transition set $S'_{I} = \{T_2, T_3\}$. Let $T_1$ has 3 actions, namely $a_1$, $a_2$, and $a_3$. Suppose $a_1$ is present in both $T_2$ and $T_3$, $a_2$ is present in only $T_3$, and $a_3$ is present in neither of $T_2$ or $T_3$. Also, assume the match score between $T_1$ and $T_2$ is 0.5 and $T_2$ and $T_3$ is 0.25. Thus, the match score will be the $max(0.5,0.25) = 0.5$ for action $a_1$, 0.25 for action $a_2$, and 0 for $a_3$. The accuracy of $S_{I}$ in terms of actions is thus $(0.5 + 0.25 + 0) / 3 = 0.25$.
\end{itemize}

\section{Cellular Specification Updates}
\label{appendix:spec-update}
The 3rd Generation Partnership Project (3GPP) introduces new generations of cellular technology, e.g., 3G, 4G, 5G, etc., roughly every 9-10 years. These new generations include a multitude of changes across all layers of the protocol stack, ranging from the physical layer (radio technology) to the upper layers (MAC, RRC, NAS, etc.) and even application layers. 
However, even after introducing a new generation, 3GPP continues to evaluate different aspects of the technology, such as security, interoperability, performance, etc. 
This leads to updates to the specifications, which can be as often as every one or two months for a document \cite{4g_nas, 4g_rrc, 5g_nas, 5g_rrc}. 
Figure \ref{fig:num-updated} shows the number of updates to two of the most important protocol layer specifications, namely RRC and NAS, in 4G and 5G for the last four years. 

Moreover, these updates to the specifications are often non-trivial. In particular, whenever a new release is introduced (roughly every year), the documents introduce significant changes ($\sim$1000 lines) for the first few versions of that release. 
Figure \ref{fig:lines-changed} shows the number of lines changed in the same set of specification documents each year. 
Here, one noticeable pattern is that 4G specifications have fewer changes than 5G, which indicates that 4G specifications are now more stable, considering it was first released 14 years ago. 
However, as new technologies and generations continue to emerge (6G and beyond), 
this trend of ever-evolving specifications is expected to persist.

\begin{figure}[t]
	\begin{center}
		\includegraphics[width=0.99\linewidth]{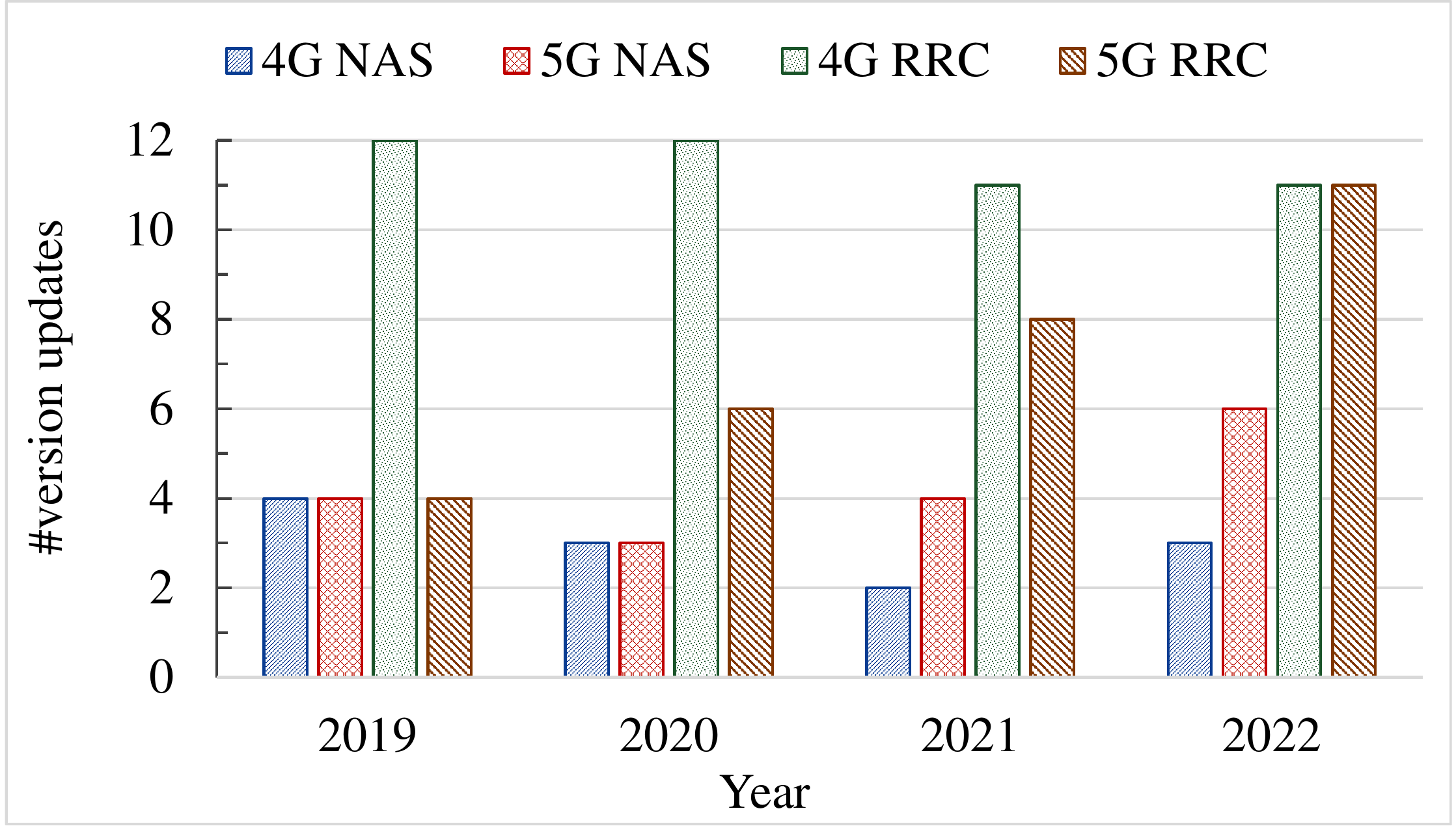}
		\caption{Number of specification updates each year.}
		\label{fig:num-updated}
	\end{center}
\end{figure}

\begin{figure}[t]
	\begin{center}
		\includegraphics[width=0.99\linewidth]{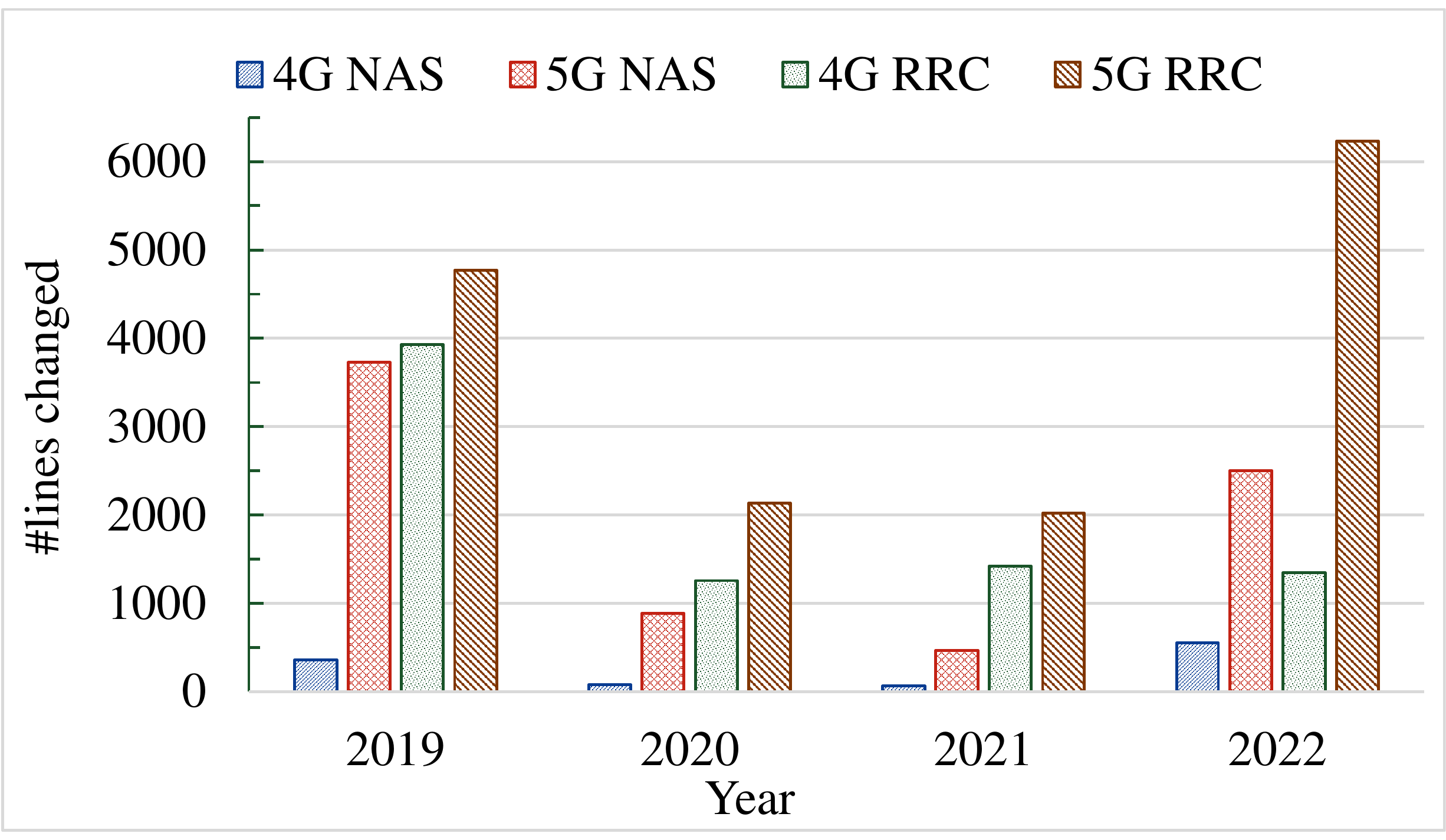}
		\caption{Number of lines changed each year.}
		\label{fig:lines-changed}
	\end{center}
\end{figure}

\section{Analyzed Procedures and Properties} 
\label{sec:properties_procedures_appendix}

\subsection{Procedures}

\begin{enumerate}
    \item Registration
    \item Deregistration
    \item Configuration update
    \item Service request
    \item Security mode control
    \item Authentication
    \item Identification
    \item GUTI reallocation
    \item Paging
    \item EMM-information
    \item RRC setup
    \item RRC security activation
    \item RRC release
    \item RRC connection reconfiguration
    \item RRC connection resume
    \item RRC connection re-establishment
    \item Measurements
    \item UE Information
\end{enumerate}

\subsection{Properties}

\subsubsection{Previous properties}

\begin{enumerate}
    \item If the base station has an RRC security context in the current state, it will exist forever unless there is a request for NAS connection establishment from a UE.
    \item If the UE requires the AS security context to be set up, the UE will eventually establish the security context.
    \item If the base station requires the AS security context to be set up, the base station will eventually establish the security context.
    \item The base station responds with a \rrcresume message if and only if the base station receives a valid \rrcresumereq message.
    \item If a UE is in the RRC-CONNECTED state and the base station sends \rrcconreconfig message, it will remain in the RRC-CONNECTED state.
    \item The replay protection assures that the same NAS message is not accepted twice by the receiver.
    \item The UE must not reset the counter while using the same security context.
    \item The AMF correctly verifies a legitimate \smcomplete message sent by the UE in response to a \smcommand message sent by the AMF.
    \item If UE initiates the service request procedure, the network will eventually perform the configuration update procedure to update the TMSI when there is no reject or equivalent message sent to the UE.
    \item If the AMF initiates the configuration update procedure, it will eventually assign a new TMSI to the device.
    \item If the AMF is in the REGISTERED state for a UE, the AMF will be in the registered state until the UE sends another registration request message.
    \item If UE initiates the registration procedure, the network will eventually connect to the network if it has a valid credential.
    \item If UE FSM is in the DEREGISTERED state, the UE will eventually move to the REGISTERED-INITIATED state, and UE authenticates MME.
    \item If the UE initiates the registration procedure, it will eventually establish the partial security context.
    \item If AMF wants to set up a security context, the UE will eventually set up/update its security context.
    \item The UE responds with an \identityresponse message only if the MME has sent an \identityreq message.
    \item The UE responds with a \smcomplete message only if and only if the UE receives a valid \smcommand message.
    \item The UE sends a \servicereq only if the MME has sent the paging message with GUTI.
    \item If the MME is in the SERVICE-INITIATED state, and UE sends \servicereq message, the MME will eventually move to the REGISTERED state.
    \item The UE will respond with the \gutireallocationcomplete message only if the MME sends \gutireallocationcommand.
\end{enumerate}

\subsubsection{New properties}

\begin{enumerate}
    \item The UE in the REGISTERED state eventually sends a \servicereq if the MME sends the paging message with GUTI.
    \item The UE does not establish a DRB before establishing the RRC security context.
    \item The UE in the REGISTERED state does not move to the DEREGISTERED state upon receiving an unauthenticated message.
    \item The UE does not accept an unauthenticated \emminfo message.
    \item The UE sends a \ueinforesponse only if the eNodeB sent the \ueinforeq message.
    \item The UE sends a \measreport with locationInfo only if the eNodeB sent the \rrcconreconfig message.
    \item The MME does not register a UE without enabling integrity protection.
\end{enumerate}



\vfill\eject

\section{Acronyms}
\begin{table}[h]
\resizebox{0.99\columnwidth}{!}{
\begin{tabular}{@{}ll@{}}
		\toprule
		\textbf{Acronyms} & \textbf{Meaning} \\
		\midrule
        \textbf{3GPP} & Third Generation Partnership Project \\
        \textbf{4G} & 4th Generation  \\
        \textbf{5G} & 5th Generation  \\
        \textbf{5G-AKA} & 5G Authentication \& Key Agreement \\
        \textbf{AMF} & Access and Mobility Management Function  \\
        \textbf{AST} & Abstract Syntax Tree \\
        \textbf{AUSF} & Authentication Server Function  \\
        \textbf{BGP} & Border Gateway Protocol \\
        \textbf{CAG} & Closed Access Group \\
        \textbf{CR} & Change Request \\
        \textbf{CRF} & Conditional Random Field \\
        \textbf{DCCP} & Datagram Congestion Control Protocol\\
        \textbf{DoS} & Denial of Service \\
        \textbf{DSL} & Domain-Specific Language \\
        \textbf{FSM} & Finite State Machine \\
        \textbf{HSS} & Home Subscriber Server \\
        \textbf{IMEI} & International Mobile Equipment Identity \\
        \textbf{IMEISV} & International Mobile Equipment Identity Software Version \\
        \textbf{IR} & Intermediate Representation \\
        \textbf{LF} & Labeled F1-metrics \\
        \textbf{LP} & Labeled Precision \\
        \textbf{LR} & Labeled Recall \\
        \textbf{LTE} & Long Term Evolution\\
        \textbf{MitM} & Man in the Middle \\
        \textbf{MME} & Mobile Management Entity\\
        \textbf{NAS} & Non-Access-Stratum \\
        \textbf{NF} & Network Function  \\
        \textbf{NG-RAN} & NG Radio Access Network \\
        \textbf{NL} & Natural Language \\
        \textbf{NLP} & Natural Language Processing \\
        \textbf{NSSAI} & Network Slice Selection Assistance Information \\
        \textbf{PDU} & Packet Data Unit \\
        \textbf{PGW} & Packet Gateway \\
        \textbf{PPTP} & Point-to-Point Tunneling Protocol \\
        \textbf{RAN} & Radio Access Network \\
        \textbf{RFC} & Request For Comments \\
        \textbf{RRC} & Radio Resource Control \\
        \textbf{SCTP} & Stream Control Transmission Protocol\\
        \textbf{SGW} & Serving Gateway \\
        \textbf{SMS} & Short Message Service \\
        \textbf{SNSSAI} & Single Network Slice Selection Assistance Information \\
        \textbf{SUPI} & Subscription Permanent Identifier \\
        \textbf{TCP} & Transmission Control Protocol \\
        \textbf{TR} & Technical Report \\
        \textbf{TS} & Technical Specification \\
        \textbf{UE} & User Equipment\\
        \textbf{UF} & Unlabeled F1-metrics \\
        \textbf{UP} & Unlabeled Precision \\
        \textbf{UR} & Unlabeled Recall \\
        \bottomrule
	\end{tabular}
    }
	\label{tab:abbreviations}
\end{table}

\end{document}